\documentclass[a4paper,12 pt]{article}
\usepackage{graphicx}
\usepackage{psfrag}
\usepackage{amssymb}
\usepackage{hyperref}
\usepackage[comma,numbers]{natbib}

\renewcommand{\d}{\mathrm{d}\mathit{}}

\newcommand{\mybibitemskiplength}{0pt} 
\def\nsim{{\tt Nsim}}
\def\nmag{{\tt Nmag}}
\def\subsubsubsection#1{{\medbreak\noindent \bf{#1}\hfill\smallskip{}}}

\usepackage{color}
\usepackage{listings}

\lstdefinestyle{defaultstyle}{}
\definecolor{lightgrey}{cmyk}{0.1,0.1,0.1,0}
\definecolor{grey}{cmyk}{0.5,0.5,0.5,0}

\begin{document}

\begin{center}
{\Large 
Continuum multi-physics modeling with scripting languages:
the {\nsim} simulation compiler prototype for
classical field theory.}
\bigbreak
{Thomas Fischbacher\footnote{e-mail: \texttt{t.fischbacher@soton.ac.uk}}},
{Hans Fangohr\footnote{e-mail: \texttt{h.fangohr@soton.ac.uk}}}\\
{School of Engineering Sciences,\\
University of Southampton\\
University Road\\
SO17 1BJ
Hampshire\\
Southampton, U.K.}
\end{center}

\paragraph{Abstract} We demonstrate that for a broad class of physical
systems that can be described using classical field theory, automated
runtime translation of the physical equations to parallelized
finite-element numerical simulation code is feasible. This allows the
implementation of multiphysics extension modules to popular scripting
languages (such as Python) that handle the complete specification of
the physical system at script level. We discuss two example
applications that utilize this framework: the micromagnetic simulation
package {\nmag} as well as a short Python script to study
morphogenesis in a reaction-diffusion model.

\tableofcontents
\section{Introduction}

The approach towards multiphysics simulations chosen for the {\nsim}
compiler prototype resolves the fundamental conflicts that arise in
multiphysics simulations in a different way than other multiphysics
packages. We hence discuss some general concepts related to continuum
multiphysics first before we give an overview over the structure of
{\nsim} and its intended uses.

\subsection{General aspects of multi-physics simulations}

When physical phenomena have to be studied quantitatively that involve
complex geometries, numerical methods that discretize space -- such as
finite element based techniques -- often are the method of choice. As
sufficient computing power to study continuum phenomena only became
available within the last few decades, continuum simulations are a
comparatively recent activity within physics. Not surprisingly, the
design of computational simulation packages more or less followed the
subject classification of physics, i.e. there are specialist packages
for electromagnetism, fluid dynamics, elastomechanics, etc. As the
simulation of different physical phenomena often requires very
different numerical techniques, this certainly is justified, but to
the physicist who is more interested in the system he would like to
study than in the technology that enables the analysis, this creates
an awkward problem: Often, interesting phenomena involve the interplay
of effects of very different nature. To a scientist or engineer who is studying
such situations in systems with geometries that are simple enough or
allow justifiable approximations so that they can be handled with pen
and paper alone, this is not much of a problem, as she has learned to
use the analytic framework to combine quantitative descriptions that
originate from different branches of physics: the discipline
sub-classification is mostly perceived as a tool to organize articles
in a library. In computer simulations of continuum phenomena, however,
this does easily become a problem, as simulation codes that deal with
different phenomena (and usually are fairly complex) are not as easily
combined as the underlying equations. This has led to the concept of
`Multi-Physics simulations' -- a curious term that is perhaps
difficult to make sense of without some understanding of this context,
in particular the artificial entrenchment of physics sub-disciplines
imposed by software requirements that only exists in such a form in
computational physics.

As `continuum multi-physics' phenomena evidently are technically very
important -- many a device works by utilizing different physical
effects, e.g. converting one form of energy into another -- this gave
rise to `Multi-Physics Simulation Packages' that have the capability
to combine continuum models from different physics sub-disciplines
into unified simulations. While the existing packages are invaluable
for the analysis of a large variety of commercially as well as
research relevant systems, their design usually is strongly governed
by the idea to support the most well established, most mature,
physical models. To the research physicist working e.g. in a field
where general consensus on the most appropriate phenomenological model
has not been reached yet (such as, for example,
micromagnetic-spintronic interactions), this results in a serious
dilemma: On the one hand, domain-specific simulation packages cannot
be used as the system to be studied involves the interplay of
different effects. On the other hand, multiphysics packages usually do
not support the very specialized sets of equations that are actively
debated in an evolving field. The only option that often remains is to
invest a large amount of programming effort to implement a new niche
software framework from scratch.

Certainly, the situation would improve if casual familiarity with
computer programming already sufficed to combine different sets of
equations as easily as a physicist is used to doing it with pen and
paper. Essentially, this means that it should be possible to specify
the physics that governs a system at the level of an accessible
high-level scripting language, as part of a simulation and data
analysis script. This idea goes beyond what many of the more generic
object-oriented multiphysics frameworks offer in that the user of the
framework does not have to have expertise in object-oriented
programming and software design. Technically speaking, the main
difference is that the ability to specify physical equations from
within a scripting language requires run-time compilation of physical
equations to fast machine code, while with an `object oriented library
framework', this translation happens in the final compilation phase
that produces an executable program. The extra degree of flexibility
offered by the run-time dynamic approach precisely parallels the
situation that arises when trying to answer the question `how can I
write a program that asks the user to enter a function and then plot
it' in a language like Python and in a language like C.

\subsection{The {\nsim} system}

This work provides a detailed report on both the conceptual as well as
implementation aspects of an automatic generic framework for the
translation of classical field theory physics to parallel numerical
simulation code. The user of the framework provides symbolic
representations of algebraic field equations as well as differential
operators which get mapped to numerical code that executes the
corresponding computations. To give an example, the discretised sparse
matrix operator that maps a scalar field $\phi(\vec x)$ to its
gradient $H_j=\frac{\delta\phi}{\delta x_j}$ gets represented by the
string ``{\ttfamily -<H(j)~||~d/dxj~phi>}''.

The simulation compiler \nsim{} parses such expressions and uses
symbolic transformations to automatically creates the appropriate fast
code (via specialization of closures) that populates the sparse finite
element operator matrices. In a parallel computation, this code is
then distributed across all nodes and used to set up the corresponding
sparse matrices in parallel. Furthermore, as the numerical solution of
ordinary differential equations for time integration often benefits
greatly from knowledge of the system's Jacobian (e.g. when dealing
with stiff ordinary differential equations), the \nsim{} compiler can
do symbolic transformations on the equations of motion (i.e. taking
derivatives) in order to auto-generate the code needed to populate the
Jacobi matrix, hence rendering the (otherwise often fairly awkward)
tasks of manually writing and debugging the corresponding low-level
code unnecessary.

\bigskip

\subsection{Architecture overview}

\begin{figure}
  \centering
  \includegraphics[width=12cm]{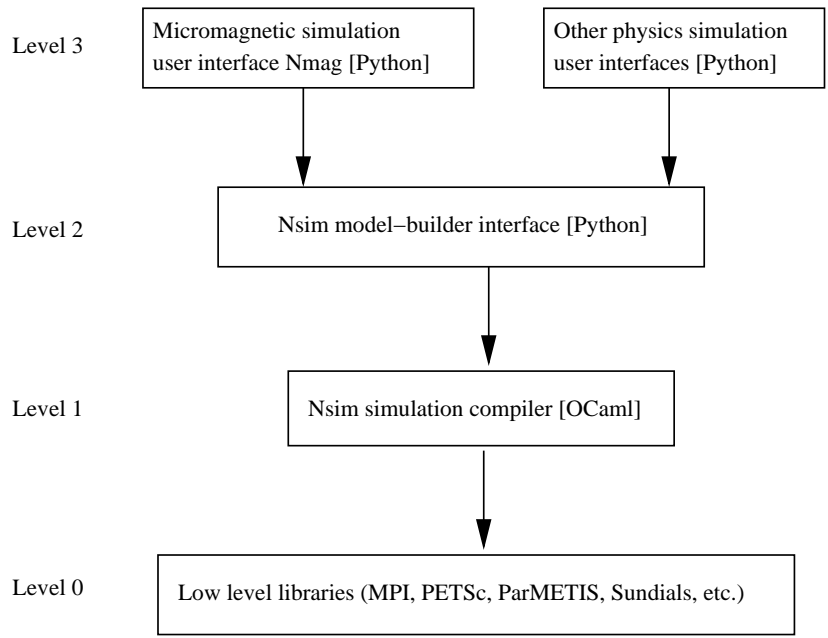}
  \caption{Overview over software layers. \label{fig:layers}}
\end{figure}

Figure~\ref{fig:layers} shows the different layers of the simulation
framework described in this work. Level~0 represents the set of
numerical libraries which provide the essential low-level
functionality for parallel simulations, the most prominent ones being
PETSc~\cite{petsc-manual2002a} for parallelizable sparse matrix linear
algebra, ParMETIS~\cite{ParMETIS} for mesh partitioning, and
Sundials~\cite{Hindmarsh2005} for integrating stiff ordinary differential
equations.  Level 1 (``\nsim{} simulation compiler'') represents the core
of the work described here, and (due to the need to do fast symbolic
transformations) has been implemented in the Objective Caml
language~\cite{OCaml}. There is a higher-level Python~\cite{Python2003a}
interface (level~2) to this {\nsim} library which can be used to
completely specify the physics to be simulated within the Python
programming language. We refer to this as the ``{\nsim} model-builder
interface'', as it will require some expert knowledge (of the {\nsim}
notation) to be able to specify a simulation at this level.

For a given research domain, such as e.g. micromagnetism, one often
observes that many groups of users (e.g. experimentalists) are mostly
interested in carrying out a limited number of simulation tasks common
to the field, which, however, often may use some
theoretical/phenomenological models which may be far from
well-established at that time. Therefore, it makes sense to implement
the functionality needed to perform such numerical simulations as an
application library on top of a generic field theory package. This
design has been chosen for the micromagnetic simulator
{\nmag}~\cite{Nmag}: The~{\nmag} package is just a thin layer of
micromagnetism-related definitions on top of the (level~2) {\nsim}
model-builder interface that provides micromagnetic capabilities in
the form of a Python extension module to the end user. The users of the
(level~3) {\nmag} system need not be aware of the lower levels, and
can easily combine micromagnetic computations with data analysis
routines offered by other, unrelated, Python packages.

We include a brief description of {\nmag} to better illustrate how the
different layers fit together, and to demonstrate their different
purposes. Figure~\ref{fig:micromag-example-code} shows an example
{\nmag} program as end-users might write it, and figure
\ref{fig:micromag-example} shows the simulation results.

\begin{figure}
  \centering
\begin{lstlisting}
import nmag
from nmag import SI

#create simulation object
sim = nmag.Simulation()

# define magnetic material
Py = nmag.MagMaterial(name = 'Py',
                      Ms = SI(1e6, 'A/m'),
                      exchange_coupling = SI(13.0e-12, 'J/m'))

# load mesh
sim.load_mesh('disk.nmesh.h5',
              [('disk', Py)],
              unit_length = SI(1e-9, 'm'))

# set initial magnetisation
sim.set_m([0,0,1])

# set external field
sim.set_H_ext([0,0,0], SI('A/m'))

# integrate over time until (meta-stable) equilibrium is found
sim.relax()
sim.save_data(fields='all') # save all fields spatially resolved
                            # -> Plot on left-hand side

sim.set_H_ext([4e4,0,0], SI('A/m'))
sim.relax()
sim.save_data(fields='all') # -> Plot on right-hand side

\end{lstlisting}
  \caption{The {\nmag} simulation script used to compute the data shown
    in figure~\ref{fig:micromag-example}. The level of computer
    programming skill required to write such short scripts that set
    up and execute basic simulations like this should be accessible to
    most research groups in an active field like micromagnetism.
    This Python program imports the {\nmag} library in the first line,
    which is the top-level user interface (level 3 in
    figure~\ref{fig:layers}) to the {\nsim} simulation package.}
  \label{fig:micromag-example-code}
\end{figure}

\begin{figure}
  \centering
  \includegraphics[width=0.49\textwidth]{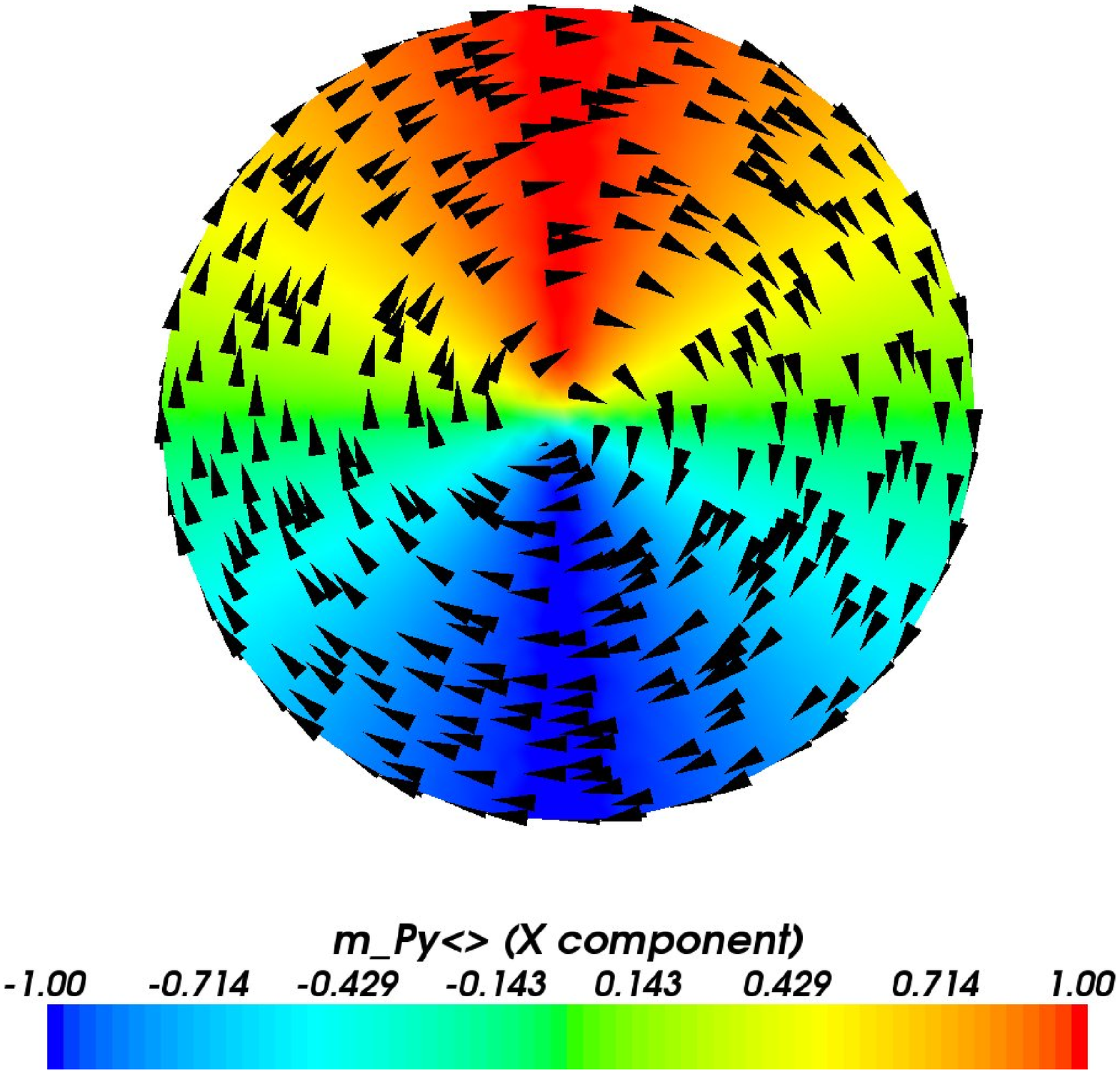}\hfill
  \includegraphics[width=0.49\textwidth]{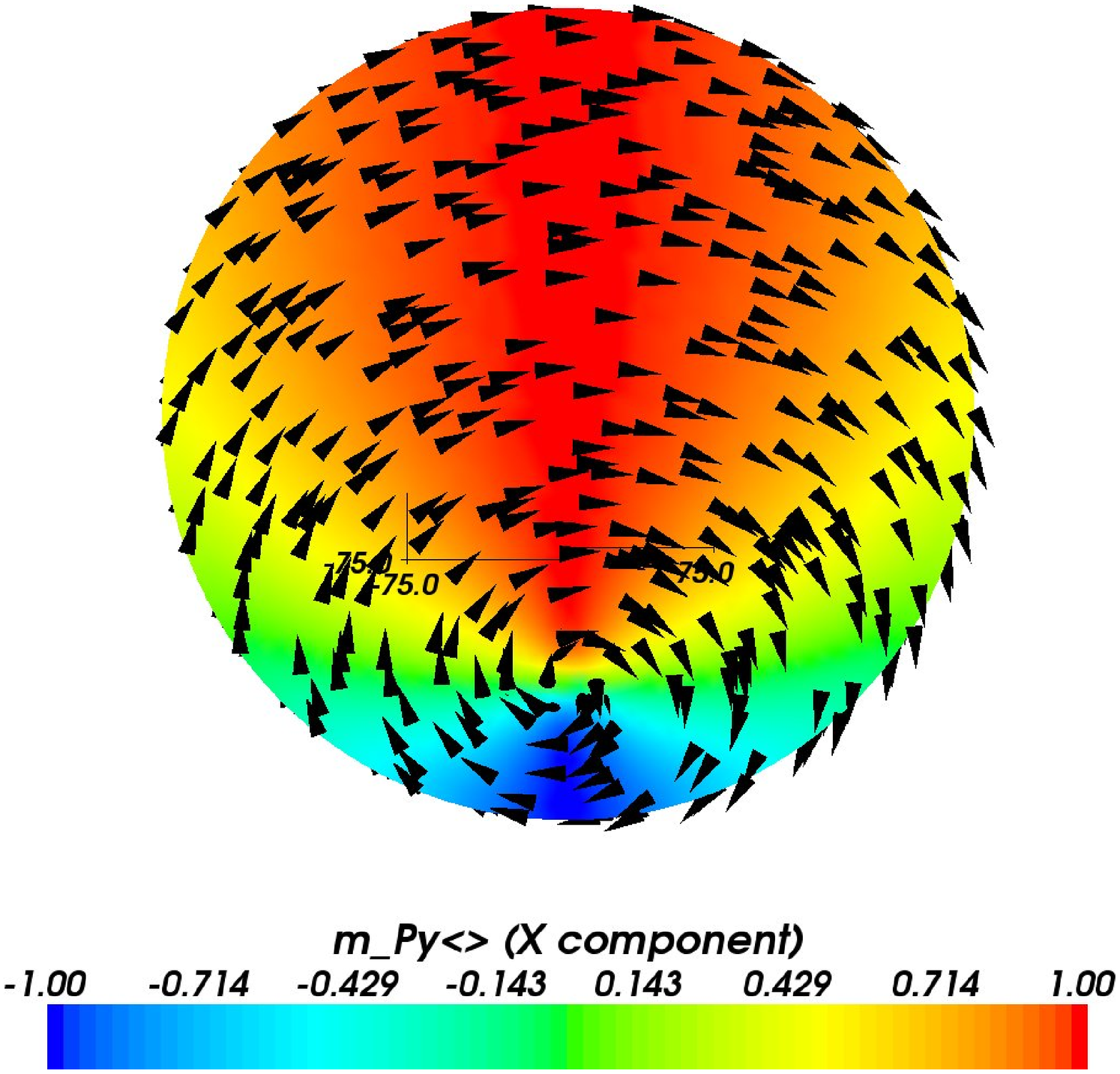}
  \caption{The results produced by the micromagnetic simulation
           script from figure~\ref{fig:micromag-example-code}}
  \label{fig:micromag-example}
\end{figure}

\subsection{Research context}

The design of this framework is driven by the need to do computer
simulations of the behaviour of engineered systems that involve novel
combinations of physical effects. As research \emph{by definition}
often has to deal with new suggestions for model equations that try to
describe physical behaviour, this situation is markedly different from
the one addressed by commercial multiphysics simulation packages that
face a strong incentive to only offer system components based on
well-established and by now widely used theoretical models. In
particular, the authors' highest priority is to provide flexibility
with regard to the adoption of novel model equations.

In more detail, the aspects addressed in this work include:

\begin{itemize}

\item Bookkeeping infrastructure for arbitrary-rank tensorial
  fields (in arbitrary dimension) associated with component-dependent
  physical characteristics (e.g. con\-ductive/{\kern0pt}non-conductive).

\item Providing a flexible, fully scriptable end user interface
  towards compiled code that can be readily combined with a large
  collection of existing software libraries for scientific analysis
  (level~2), and may be used to quickly put together domain specific
  simulation packages (level~3).

\item Automatically producing fast machine code from a symbolic
  representation of (potentially non-linear) physical equations that
  can be changed at run time (e.g. interactively).

\item Embedding symbolic term transformations in a numerical
  simulation framework in order to automatically generate the Jacobian
  needed for the time integration of stiff systems.

\item Relieving the user from having to deal with memory management
  issues.

\item Psychological aspects related to how convenient different
  syntactic approaches towards specifying boundary conditions,
  equations of motion, etc. turn out to be in real use scenarios.

\end{itemize}

%Major issues that also eventually have to be addressed but are not
%covered in this work include:
%
%\begin{itemize}
%
%\item Specific discretisation strategies for physical systems
%      (such as adaptive meshing, finite volume methods, etc.).
%
%\item Estimating numerical errors.
%
%\item Profiling parallel execution.
%
%\end{itemize}

This work describes in detail the \emph{prototype} (level~1)
implementation of the {\nsim} simulation compiler at the time of this
writing. This compiler forms the basis for the multiphysics
micromagnetic simulation library {\nmag} (level~3). As some parts of
{\nsim} still experience considerable evolutionary changes, this work
inevitably suffers from a `moving target' problem in the sense that
there are some loose ends that will be addressed in the future, and
that the ongoing development is driven by the input from the wider
academic community.

The only spatial discretisation scheme supported by {\nsim} is finite
element discretisation using unstructured simplicial meshes (i.e.
tetrahedral meshes in three dimensions). While the structure of the
system would support generalizations towards other methods for
discretising space, this has not been implemented so far.

\bigskip

The outline of this work is as follows: Section~\ref{sec:motivation}
describes the phenomenological model that gave the original motivation
for this work -- micromagnetism -- and some aspects of both the
relevant physics and its implementation. Section~\ref{sec:background}
provides the context for the simulation compiler presented here.
Section~\ref{sec:form-spec-phys} explains key aspects of the
model-builder interface. Section~\ref{sec:nsim-comp-prot} then
explains in detail what low level techniques are used in the prototype
implementation to provide this functionality.
Section~\ref{sec:limitations} comments on limitations, before we
summarize in section~\ref{sec:conclusion-outlook}.
Appendix~\ref{sec:appendix} provides a complete worked out (level~2)
Python example that shows how to use {\nsim} to set up a continuum
physics simulation from scratch.

\section{Motivation/Example}\label{sec:motivation}

\noindent The scientific incentive for this work was provided by the
need to implement an extensible simulation framework for computational
micromagnetism that is easy to extend towards novel combinations of
effects, e.g. to quantitatively study magnetoelectric, magnetothermal,
and other `magnetism+X' phenomena. We discuss the structure of the
micromagnetic {\nsim} application, {\nmag}, in some detail in order to
give an example of the techniques and approaches available to
implementors of libraries that deal with other systems.

\subsection{Micromagnetism}
\label{sec:micromagnetism}

\noindent As mesoscale magnetism involves the interplay of a number of
physical effects of very different nature yet similar strength,
simulations of mesoscopic magnetization dynamics are at the same time
somewhat challenging to set up (due to the amount of effort required
to deal with each relevant effect separately), yet rich in interesting
behaviour. It is precisely this richness in structure caused by the
competition between different physical effects that makes magnetism an
interesting candidate for a number of applications, first and foremost
data storage utilizing hysteretic effects and multiple (thermally
sufficiently stable) local energetic minimum configurations.

The selection of effects supported by a micromagnetic simulation
framework is to some degree a matter of practicability and community
consensus. From a condensed matter perspective, magnetism interacts
with an enormously broad range of other solid state physics sectors
through thermomagnetic, magnetoelastic, magnetoelectric and other
effects. As a number of these offer interesting potential for novel
applications: for example, new magnetic media storage devices might
employ heating to above the ferromagnetic phase transition temperature
in Heat Assisted Magnetic Recording (HAMR)~\cite{Rottmayer2006} or
exploit spin-polarized electric currents in novel 3d-magnetic memory
(e.g. ``Racetrack Memory''~\cite{Parkin2008}) devices. Therefore, one
would like a micromagnetic simulation framework to be extensible
enough to easily accommodate such specific additional effects. If it
can be designed in such a way that extensions are treated on the same
footing as already existing features, then this clearly should be
regarded as an important advantage.

From the perspective of using a specific challenge as guidance when
devising a versatile general framework, it is important to choose a
primary application that is sufficiently rich in structure to cover
and test the majority of generic features required. It hence seems
reasonable to expect that most problems of practical relevance to
which the {\nsim} framework can be applied are structurally much
simpler than micromagnetism.

We briefly describe the structure of the field theoretical model of
micromagnetism as originally presented by Brown~\cite{Brown1963a}.

The fundamental degrees of freedom are provided by a vector field
$\vec M(\vec x)$ describing the local magnetization, more precisely,
the local magnetic dipole momentum density: When integrated over a
region~$\Omega$ of space, $\int_\Omega \vec M(\vec x) d^3\vec x$ gives
the total magnetic dipole momentum contained in that region.

Physically, the magnitude of $\vec M$ is determined by local material
properties, hence constant. (At least in the commonly used
zero-temperature micromagnetic model). Thus, there actually are only
two dynamical degrees of freedom at every point -- the direction of
the magnetization vector in space.\footnote{In group theory language
  one may think of this as a $S^2 \simeq SO(3)/SO(2)$ coset model.} In
simulations, one hence usually prefers to take a unit-normalized field
$\vec{m}={\vec{M}}/{|\vec{M}|}$ as the fundamental dynamic quantity.

The dynamics of the field $\vec M(\vec x)$ is governed by the
so-called Landau Lifshitz and Gilbert (LLG) equation which describes the
reaction of magnetization to an effective magnetic field. This
equation is local and non-linear, but still polynomial in the fields
involved. As $|\vec M|$ is constant, $d\vec M/dt$ must be
perpendicular to $\vec M$, and in the simplest possible
phenomenological model should be proportional to the effective
magnetic field strength experienced locally. So, the most general
expression possible is:
\begin{equation}
\label{LLG}
\frac{\d}{\d t}\,\vec M(\vec x) =
 c_1 \vec M(\vec x) \times \vec H_{\rm eff}(\vec x)
+ c_2 \vec M(\vec x) \times\left(\vec M(\vec x) \times \vec H_{\rm eff}(\vec x)\right),
\end{equation}
which is just the LLG equation. The $\vec M\times\vec H_{\rm eff}$
term gives rise to magnetic precession, as is easily seen by noting
that this is perpendicular to both~$\vec M$ and~$\vec H_{\rm eff}$,
hence does not change the angle between $\vec M$ and $\vec H_{\rm
  eff}$. Precession usually dominates the motion of a magnetic moment
in an external field. The other term is dissipative in nature and
causes the magnetization to eventually align with the applied field.

In more systematic tensor-index notation (where we now also suppress
the spatial dependency on $\vec x$ and use Einstein's summation
convention), this can be written as:
\begin{equation}
\label{LLGtensor}
\frac{\d}{\d t}\, M_i =
 c_1 \epsilon_{ijk} M_j H_{{\rm eff}\, k}
 + c_2 \epsilon_{ijk} M_j \epsilon_{kmn} M_m H_{{\rm eff},n},
\end{equation}
with~$\epsilon_{ijk}$ being the fully anti-symmetric rank-3 tensor
that is defined by:
\begin{equation}
\det(\vec u,\vec v,\vec w) = \epsilon_{abc} u_a v_b w_c.
\end{equation}

While this equation automatically ensures the conservation of $|\vec
M|$, in numerical simulations that use three coordinates to encode the
direction of $\vec M$, rounding may cause some drift that violates
this constraint. One convenient way to deal with this problem is to
modify the equation of motion by adding another term which physically
always is zero, but keeps numerical drift in check:
\begin{equation}
\frac{\d}{\d t}\, M_i =
 c_1 \epsilon_{ijk} M_j H_{{\rm eff}\, k}
 + c_2 \epsilon_{ijk} M_j \epsilon_{kmn} M_m H^{({\rm eff})}_n
 + c_3 M_i \left(M^2-M_k M_k\right).
\end{equation}

The {\nsim} prototype exclusively employs tensor-index notation, as
this is both systematic enough to be processed easily yet powerful
enough to express even uncommon equations, e.g. involving angular
momentum flow.

Contributions to the effective magnetic field strength $\vec H_{\rm
  eff}$ include the externally applied field $\vec H_{\rm ext}$ (which
may be Earth's magnetic field, or a field applied in a laboratory
setting), the so-called `demagnetization field' $\vec H_{\rm demag}$
that gives the magnetic field generated in one location by the
combined effects of the magnetic dipoles distributed throughout the
system, as well as others that, for the purpose of magnetization
dynamics, behave like contributions to the effective magnetic field,
but are not linked to a true magnetic flux density. One of these is
the effective magnetocrystalline anisotropy~$\vec H_{\rm anis}$: In
crystals, alignment of the direction of magnetization with a
crystallographic axis may be energetically favorable, and a magnetic
moment starting out in a direction of non-minimal energy will precess
and ring down in a way that is just equivalent to the dynamics modeled
by the LLG equation with a direction-dependent effective `equivalent'
magnetic field strength. Physically speaking, the magnetocrystalline
anisotropy arises from the coupling of spin and orbital angular
momentum of the magnetic electrons, hence is a quantum effect. Another
related quantum effect gives rise to a further `magnetic field
equivalent' contribution: The quantum mechanical overlap between
electron orbitals gives rise to an `exchange field strength' $\vec
H_{\rm exch}$ which normally tries to align magnetic moments in
parallel that are close to one another. Usually, this is the most
important contribution to the effective field strength.

The demagnetizing field $\vec H_{\rm demag}$ is computationally most
expensive due to its `every individual magnetic moment interacts with
every other magnetic moment from the rest of the sample' nature. This
is of the structure:
\begin{equation}
H_{{\rm demag},j}(\vec x) = \int d^3y\,G_{jk}(\vec x,\vec y)\, M_k(\vec y)
\end{equation}
where the kernel $G_{jk}(\vec x,\vec y)$ gives the magnetic field at 
position $\vec x$ produced by a localized Dirac dipole at $\vec y$.

The effective exchange field strength is phenomenologically modeled by
taking the vector (i.e. component-wise) Laplacian of the
magnetization:
\begin{equation}
H_{{\rm exch},j}(\vec x) = c_{\rm exch}\, \frac{\partial}{\partial x_k}\,\frac{\partial}{\partial x_k} M_j(\vec x).
\end{equation}

The magnetocrystalline {\em anisotropy} term usually is described
through a truncated polynomial expansion, which is a {\em local}
function of the magnetization $\vec M(\vec x)$, or, in tensor-index
notation:
\begin{equation}
H_{{\rm anis}\,j}= \delta E[\vec M]/\delta M_j
\end{equation}
with the energy function being of the form
\begin{equation}
E[\vec M]=E^{(0)} + E^{(1)}_j M_j + E^{(2)}_{jk} M_j M_k + E^{(3)}_{jkl} M_j M_k M_l + \ldots
\end{equation}

(Physically speaking, $E^{(0)}$ is irrelevant and $E^{(1)}_j$ would
behave like another contribution to the externally applied field.
Nevertheless, this is the most general form.)

\subsection{A micromagnetic simulation}\label{sec:example-nmag}

\noindent
Some results of a simple micromagnetic example calculation are shown
in figure~\ref{fig:micromag-example}, and have been computed using the
script shown in figure~\ref{fig:micromag-example-code}.

The system simulated is a magnetic disk with a diameter of
150~nanometers~(nm) and a thickness of $5\,\rm nm$. The material
parameters are chosen as $|\vec M|=10^6\,\mathrm{A/m}$, and $c_{\rm
  exch}=1.3\cdot 10^{-11}\,\mathrm{J/m}$, roughly similar to those of
PermAlloy.

The local exchange coupling favors parallel alignment of the
magnetization $\vec{M}$ over short distances, as the black cones
representing the direction of magnetization in
figure~\ref{fig:micromag-example} show. This interaction competes with
the demagnetization interaction which favors a configuration where
magnetic North and South poles connect and which minimizes surface
charges. The physical solution for the given geometry and material
parameters in the absence of any external magnetic fields is shown in
figure~\ref{fig:micromag-example} (left) and represents a vortex. This
closed flux configuration of the magnetization minimizes the
demagnetization energy, and incurs only a small energy penalty from
the curvature in the magnetization.

Figure~\ref{fig:micromag-example} (right) shows the resulting
equilibrium configuration if an magnetic field of
$H_\mathrm{ext}=40\cdot 10^3\,\mathrm{A/m}$ is applied, pointing from
the left to the right. Comparing the situation without an applied
field shown in figure \ref{fig:micromag-example} (left) to the
situation with the applied field (right), the center of the vortex
moves downward. This minimizes the overall energy of the system
because a large fraction of the magnetization aligns with the external
field.

This is a fairly simple micromagnetic example which should convey the
kind of complex physics emerging from competing interactions. Problems
like this can often (especially in complicated geometries) only be
tackled successfully using numerical simulations.

\subsection{Multi-physics micromagnetism}
\label{sec:multi-phys-micr}

The effective field $
\vec H_{\rm eff}(\vec x)=
 \vec H_{\rm ext}(\vec x) 
+\vec H_{\rm demag}(\vec x)
+\vec H_{\rm exch}(\vec x)
+\vec H_{\rm anis}(\vec x),
$ collects the contributions outlined above (which all are of computationally
very different type) and make up the conventional micromagnetic model.

There is growing need to include additional physics to these
simulations, such as e.g. the extra torque on magnetic moments caused
by (spin-polarized) electric current densities. This is required to
simulate the effects that may be employed in the data storage and
processing devices of the future.

In micromagnetism, most additional contributions to the equation of
motion can be expressed through local polynomials in other fields,
spatial differential operators of order at most two, and integrals
over propagators from long-range interactions. There are reasons based
on insights from renormalization theory on the general structure of a
hierarchy of short-range corrections to classical field theory that
suggest that this is a fairly common situation that should arise in
most other phenomenological models of mesoscopic physics as well.
Essentially, these concepts are applicable in most applications where
the system's free energy functional can be given as a truncated series
expansion.

\section{Multi-physics simulations}\label{sec:background}

\noindent
The need for and growing importance of multi-physics simulations has
been recognized in the computational science and engineering community
and industry for some time~\citep{Parry2000a,Tartakovsky2005ashort}. 

Multi-physics simulations are a relatively new computational field in
materials science, and often simulations are carried out by using two
(or more) simulation codes that simulate one phenomenon each. The
different codes are run subsequently and iteratively until a
self-consistent solution has been found. However, in general it is
desirable to solve all the equations simultaneously (even though this
often means existing codes cannot be used) so that -- in particular --
time dependent processes can be studied more easily.

Particular sectors of computational engineering have been using
multi-physics simulations for a long time, including fluid-structure
interactions and aeroelasticity, and combustion chamber simulations.

Commercial packages have become available for mainstream engineering
applications~\citep{Physica2006a,Comsol,ANSYS}
but there is still need for tools with either specialized
functionality for niche or emerging research domains or capability to
exploit particular hardware for high performance computing.

In more detail, commercial multi-physics simulation tools (such as
Comsol~\citep{Comsol} (formerly FEMLAB), and ANSYS~\citep{ANSYS}) can
perform some of the computations required for micromagnetic
simulations (such as computing the current density distribution from a
given applied voltage, or computing the diffusion of heat). However,
they (at the time of this writing) cannot carry out the complex
high-frequency micromagnetic and spintronic dynamics.

While {\nsim} -- to the authors' best knowledge -- is presently unique
in its combined support for runtime translation of model equations,
transparent parallelization, generic tensor-index symbolic notation
for physical equations, and end user flexibility through utilization
of the capabilities of a well-established scripting language, other
frameworks exist that follow an approach towards field theory
simulations which has some overlap with {\nsim}'s. These include the
commercial packages listed above, Diffpack~\cite{Diffpack},
Fastflo~\cite{FastFlo}, as well as the free systems
Elmer~\cite{Elmer}, FEniCS~\cite{Fenics}, FreeFem~\cite{FreeFem},
OOMPH-Lib~\cite{OOMPH-LIB}, OpenFEM \cite{OpenFEM},
OpenFOAM~\cite{OpenFOAM}, and the system presented
in~\cite{Abercrombie2003}.

\section{Formal specifications of physical systems}\label{sec:form-spec-phys}

Ideally, automated compilation of field theoretic systems should be
able to proceed from a purely formal specification of the equations of
motion, the geometry of the system, and the initial and boundary
conditions to a working simulation of the time evolution of that
system. Work on the \nsim{} prototype system showed major differences
in the usability and practicality of different notational conventions.
At the time of this writing the symbolic conventions suggested must be
considered as still being in a state of flux. We hence report our
present findings, warning the reader that these are preliminary and
possibly subject to some degree of future change as our understanding
of the psychological aspects of mapping physical equations to the
ASCII character set increases.

\subsection{A perspective on abstraction}

Considering the history of computer programming, an evolutionary step
that has come into reach now is to provide and utilize the linguistic
capabilities to directly express field theory concepts and ideas in
computer code and leave the details of translating these to (possibly
parallel) matrix/vector operations to a compiler. While in the
previous visible evolutionary trend towards ever higher forms of
abstraction, high hopes and big dreams that ultimately failed to
manifest were associated with every step, important progress
unquestionably has happened nevertheless. While at no point, we
reached the stage that would `render debugging unnecessary', a higher
level of abstraction allows one to spend effort more efficiently on
dealing with aspects that are closer to the properties of the part of
the real world one tries to model than to the characteristics of the
computing machinery on which this model is implemented.

\subsection{Fundamental concepts}

A framework designed for the purpose of simulating multiple
interacting physical sectors (e.g. micromagnetism and heat conduction)
requires some abstract notions. It should be pointed out that both
these concepts \emph{as well as their implementation in the 
\nsim{} prototype}\footnote{With the exception of some surface/surface 
interaction integrals that use properties specific to three
dimensions} are sufficiently general to support physics in arbitrary
dimension, rather than just three-dimensional space. We use the
following terms:

\begin{itemize}

\item A {\em region} is a (potentially disconnected, potentially
  multiply connected) subset of the cells of a mesh that describe a
  material with uniform physical properties. Regions are numbered~1,
  2, \dots and may have additional \emph{attributes} such as `{\tt
    magnetic}' or `{\tt conducting}' that then can be used when
  specifying differential operators. As this notion is especially
  important to define boundary conditions, we extend it to include
  pseudo-regions associated to non-meshed outer space, which carry
  negative numbers~-1, -2, \dots This allows us to use pairs of
  integers to describe boundaries and designate different types of
  boundary conditions on different parts of the outer boundary of a
  material -- e.g. Dirichlet (constant potential) boundary conditions
  on some faces of a conductor and and von Neumann (no current flowing
  off the sample) boundary conditions on others.

\item A {\em site} is a position in the mesh that can carry degrees of
  freedom. When using first order elements, these are just the
  vertices of the mesh. When using second order elements, this would
  also include edge midpoints.

\item An {\em element} is a prescription to associate degrees of
  freedom to a cell of a mesh (e.g. a magnetization vector and a
  conductivity scalar). Usually, there will be one element type for
  every region in the mesh.

\item A {\em degree of freedom} is a single (real floatingpoint)
  number associated to a site that has a {\em name} which consists of
  a string -- the {\em stem} -- and an array of integers (possibly of
  length zero) that describe its {\em indices}, where index counting
  starts at zero. The name of the degree of freedom describing the
  x-directed component of the electrical field strength hence would be
  {\tt E[0]}, while the name of the z-directed flow of x-directed
  momentum could be named {\tt J\_p[0,2]}.

\item A {\em field} is a collection of degrees of freedom obtained by
  associating elements to the cells of a mesh.

\item A {\em subfield} is a subset of the degrees of freedom of a
  field that share the same name stem. The rationale underlying this
  concept is that we may have to collect multiple physical fields of
  different origin and nature into a single entity. For example, in a
  multi-material micromagnetic simulation, sites belonging to the
  boundary separating two magnetic materials will have to carry two
  different types of magnetization.

\item A {\em primary field} is a field that describes those aspects of
  the system that can in principle be influenced by the experimentor.
  In micromagnetism, both magnetization and externally applied field
  strength are primary fields.

\item An {\em auxiliary field} is an intermediate quantity derived
  either directly from the primary fields, or from other auxiliary
  fields. In micromagnetism, the effective magnetic field strength
  $\vec H_{\rm eff}$ is an auxiliary field.

\end{itemize}

\subsection{The language}\label{sec:language}

The formal specification of a physical system to be simulated consists of:

\begin{itemize}

\item A description of the geometry (including information about regions).

\item A description of the fields associated to (different regions in) 
the system.

\item The specification of initial and boundary conditions.

\item A collection of (local and differential) equations that determine 
how auxiliary fields are obtained from primary fields.

\item (optionally) tweaking parameters for improving the computational 
efficiency of some particular parts of the simulation.

\end{itemize}

Considering computer simulations in the broader context of their
scientific and industrial environment, there may be important
extrinsic aspects associated to some pieces of such a description: In
particular, while it may be desirable to be able to define the
geometry of some sample through a small number of equations, we can
realistically expect that many users will also want to utilize
pre-existing geometry descriptions e.g. produced by some mesh
generating CAD software. While the \nsim{} prototype includes some
basic mesh-generating capabilities to demonstrate the concept of fully
equation-based analytic system specifications~\cite{BordignonPhD}, 
this so far has not been evolved into a highly efficient mesh generating 
subsystem. As externally generated meshes can be utilized, work on the 
geometry specification aspects is presently regarded a lower priority issue.

In the \nsim{} prototype, the layout of a field is specified by
providing the association of elements to mesh regions. An element is
either a primary element, associating a complete set of tensor degrees
of freedom (such as $E_x, E_y, E_z$) to each site in a mesh cell, or
is a fusion of other elements. Occasionally, we want to work with a
collection of fields of identical layout (i.e. component-wise addition
of the entries of the numerical arrays is a sensible operation) in
which the degrees of freedom are nevertheless addressed through
different names. In the micromagnetic application \nmag{}, this
situation occurs e.g. when contributions to the total effective
magnetic field strength {\tt H\_{eff}} which are of very different
physical origin are summed.  Experiments with a number of different
strategies have shown that a convenient and uniform approach towards
such a situation is to introduce `derivatives' of existing field
layouts which are produced by renaming the name stems of the degrees
of freedom. A code example that demonstrates this technique can be
found in appendix~\ref{sec:appendix}.

\subsubsection{Differential operators}

Differential operators, which get mapped to sparsely occupied matrices
acting on the data arrays of fields, are specified as strings that get
parsed according to a grammar which is inspired by quantum mechanical
bra-ket notation and also supports the Einstein summation
convention. In this notation, the Galerkin discretisation of e.g. the
heat diffusion operator
\begin{equation}
M_{ij} = - \int_{\mathcal V} \frac{\partial}{\partial x^k} \phi^{(DT)}_i \frac{\partial}{\partial x^k} \phi^{(T)}_j d{\mathcal V}
\end{equation}
is given as:
\begin{equation}
\mbox{\tt - <d/dxk DT || d/dxk T>, k:3}
\end{equation}
while the operator computing the divergence of a tensor-valued field 
of rank two could be specified as:
\begin{equation}
\mbox{\tt <divJ(a) || d/dxb J(a,b)>, b:3}
\end{equation}

While this `core notation' has been found to be both viable and
convenient to use, some of the more complicated extensions are still
in a state of flux, especially when involving boundary conditions. We
briefly describe each presently supported extension in turn as well as
hint at a context in which these become relevant.

\subsubsubsection{Surface derivatives}

While the \nsim{} prototype at present is based on simplicial (in 3d,
tetrahedral) rather than singular homology, i.e. does not allow the
association of degrees of freedom to sub-simplices of volume simplices
(such as point or line charges), surface derivatives play such an
important role for continuity equations that a special notation has
been introduced to deal with them. When sparse operator matrices are
being computed from abstract operator descriptions by working out
integrals over cells of products of shape functions, the field prefix
{\tt d/dxj} causes the replacement of the corresponding field basis
function by its coordinate derivative. Likewise, the field prefix {\tt
D/Dxj} effects the modifications required to produce matrix
contributions that correspond to surface divergence jumps for simplex
faces across which the field ceases to exist. (The volume integral
over the simplex is replaced by surface integrals over the relevant
faces, multiplied by the $j$-th component of the surface `normal'
(rather, $(d-1)$-vector).  The translation of expressions in which two
surface derivatives occur in the same bra-ket, such as {\tt <D/Dxj phi
|| D/Dxj phi>}, causes an error, as such a product of distributions is
not well defined mathematically.)

In micromagnetism, these surface derivatives are required to correctly
compute the surface divergences of the magnetization field which
enter the computation of the demagnetizing field as source
terms. (Note that for homogeneously magnetized systems, these surface
magnetic charges are the only sources of the demagnetizing field.)

\medbreak

\subsubsubsection{Region and boundary restrictions}

We may want to restrict some contributions to a differential operator
to degrees of freedom associated with some particular mesh regions, or
region boundaries. This can be done by adding a restricting qualifier
in angle brackets to the name of the degree of freedom. Experience
with earlier versions of the {\nsim} prototype has shown that it is
surprisingly tricky to define a region specification language that is
both conceptually simple and at the same time flexible enough in order
to support all the common use cases. The present version of {\nsim}
uses a simple grammar for boolean logic that refers to region
properties associated with materials and allows specifications such as
{\tt J[not magnetic](2)} or {\tt J[conducting and insulated\_boundary](1)},
and could also handle more complicated cases such as 
{\tt T[(ferroelectric or ferromagnetic) and not boundary]}.
While this is not expected to be the final word on region 
specifications\footnote{especially in conjunction with code-generated
equations where an infix notation of logic operations may not be
the most natural and convenient representation}, more simplistic
approaches have been shown to be too clumsy in practice.

\subsubsubsection{Matrix shortening}

When employing region-based restrictions, we may want to construct a
`shortened' version of some matrix operator which maps (potentially)
shortened data vectors to (potentially) shortened data vectors.
Shortened vectors only contain the degrees of freedom selected by some
particular region restriction and play an important role e.g. when
providing boundary values for (say, Dirichlet) boundary conditions, or
exchanging data between the FEM and BEM parts of some hybrid FEM/BEM
code. The syntax presently used in the \nsim{} prototype is to append
a semicolon-separated qualifier to an operator specification that
gives the left- and right-hand-side restrictions on the degrees of
freedom, as in:
\[
\mbox{\tt <phi || phi>; (L||R)=(*||phi[boundary])}.
\]

\subsubsubsection{Residual gauge fixing}

When using a linear solvers to invert discretised operators, such as
when solving a Laplace equation to obtain a potential from a charge
density, we will have to address the `residual gauge symmetry' that
corresponds to a (spatially constant) shift of the potential $\Delta
\psi = \Delta (\psi+c)$ which reduces the rank of the Laplace
operator's matrix by one for every component of connection. Linear
algebra libraries usually provide functions to deal with such
situations\footnote{such as the {\tt MatSetNullspace} function in the
PETSc~\cite{petsc-manual2002a} library employed by \nsim{}} and generally
should be preferred over other solutions (such as introducing local
modifications to the matrices), as other approaches may cause a
noticeable slowdown due to unnecessary communication overhead in
parallel computations. At the time of this writing, information about
such null spaces has to be specified in {\nsim} when setting up a
linear solver to invert a differential operator. In earlier versions,
this was provided as a grammar extension to the operator specification
itself. It is not yet entirely clear what the best approach is here.

\subsubsubsection{Dirichlet boundary conditions}

When working with Dirichlet Boundary Conditions, we may want to
exclude certain rows and columns from the matrix but introduce ones on
the diagonal in order to maintain regularity. This at present can be
specified using semicolon-separated modifiers of the form {\tt LEFT=RIGHT}, 
as in:
\[\mbox{{\tt <d/dxk phi[vol] || d/dxk phi[vol]>; phi[boundary]=phi[boundary], k:3}},\]
but again, this notation is considered to be preliminary and somewhat
unsatisfactory.

\subsubsubsection{Periodic boundary conditions}

There are a number of subtle aspects associated to periodic boundary
conditions which are explained in detail in~\cite{Macrogeometry}.
Essentially, periodicity is specified by associating the mesh with
information about sites that are to be identified (which need not be
at outer boundaries). Linear operators discretized on such meshes
usually will want to respect periodicity. In \nsim{}, degrees of
freedom subject to periodic identification occur multiple times in
field data vectors, as this greatly simplifies some operations, such
as interpolating field values over the mesh. Duplication of degrees of
freedom must not occur, however, in the state vectors on which the
time integrator acts, as this would cause serious problems with stiff
solvers. Therefore, \nsim{}'s support for periodic boundary conditions
requires mapping data vectors back and forth between `long'
representations with duplicate entries and `short' representations
without, as well as special tricks when automatically generating a
Jacobi matrix that acts on `short' vectors but may involve (in its
definition) differential operators that deal with `long' vectors.

As a consequence of this approach, operators at present have to use a
periodicity modificator, as in {\tt <phi||phi>; periodic=phi}. The
effect is that matrix contributions computed for one instance of a
degree of freedom subject to periodic identification will be added for
other copies as well.

\subsubsubsection{Generalized second order operators}

Generalized forms of Laplace's operator play an important role in many
systems. One prominent example is the computation of the electric potential
from fixed surface potentials when conductivity is a function of space 
(or indeed even a dynamic field). Here, we have 
\begin{equation}
\vec j=\sigma E,\quad E=-\vec\nabla \Phi,\quad \vec \nabla\cdot\vec j=0,
\end{equation}
hence we usually want to solve an equation of the form
\begin{equation}
\vec\nabla\left(\sigma\vec\nabla\Phi\right)=0
\end{equation}
with given boundary conditions. In a numerical discretisation, there
is a subtle difference between applying two first-order operators in
succession, and applying one second-order operator instead that
captures the same physical idea: This is easily seen by observing that
the split operator approach would produce next-to-next-neighbor
contributions where the single operator approach would only have
next-neighbor contributions.

The syntax used in \nsim{} to specify operator matrices that depend on
such `middle fields' is {\tt <d/dxk rho|sigma|d/dxk phi>,k:3}. In the
version of {\nsim} published at the time of this writing, a working
implementation of code for populating and recomputing sparse operator
matrices that involve middle fields is present, but this functionality
is not yet accessible from the user level.

\subsubsection{Local non-linear equations}

Apart from auxiliary fields that are computed by applying differential
operators to more fundamental fields, and hence `peek' the physics in
the neighborhood of a given site, there also are fields which are
being obtained directly from information given at the site in
question: These fields are usually called `algebraic'. In
micromagnetism, we encounter both situations in different
contributions to the effective total field strength {\tt H\_total}:
The exchange field strength is obtained by applying a differential
operator to the magnetization field that measures how much it `bends'
in the neighborhood of a given point, while the anisotropy field
strength is a (potentially complicated) polynomial in the local
components of the magnetization vector only.

Any formalism devised to express the computation of auxiliary fields
from other fields residing at the same site imposes a restriction on
the versatility of the framework unless it is computationally
complete. Hence, the \nsim{} prototype provides both a convenient
tensor-index formula notation for local operators as well as a way to
by-pass it and directly provide a piece of C code that can implement
arbitrarily complicated local transformations on fields. Internally,
the tensor notation mechanism sits on top of the more flexible
mechanism that allows providing C code and uses both a term parser and
C code generator. In both cases, the C code that implements local
operations utilizes a number of machine-generated macros that check
the presence of particular subfields at a given site and allow access
to their contents.

The problem with user-provided C code is that it presently cannot be
derived symbolically to auto-generate entries in the Jacobi matrix
required by the ODE solver for time evolution, so systems that feature
both local equations not expressible in the present grammar and stiff
behaviour that requires obtaining a good approximation to the Jacobi
matrix are not well supported yet. The general expectation is that in
such situations, one can either afford the extra effort required by
the time integrator library to internally generate an approximation to
the Jacobian, or ask for an extension of the \nsim{} prototype to
provide additional expressive power to the grammar of local
operations: The \nsim{} developers would like to hear from users who
run into such problems.

The formula notation based specification of a local operation must
provide the ranges of silent indices, a listing of auxiliary local
buffers to be used intermediately in the computation, and a sequence
of assignments whose left hand side is a subfield degree of freedom
and whose right hand side is a tensor polynomial. This tensor
polynomial may involve pre-defined constant tensors such as the
$\epsilon_{ijk}$, which is used to express vector cross products. The
Landau Lifshitz and Gilbert equation of micromagnetism~(\ref{LLG}) can
be expressed in this notation as:

\bigskip

\begin{minipage}{10cm}
\verb$%range i:3, j:3, k:3, p:3, q:3;$\\
\\
\verb$dmdt(i) <-  C1 * eps(i,j,k) * m(j) * H_eff(k)$\\
\verb$           +C2 * eps(i,j,k) * m(j) * eps(k,p,q) * m(p) * H_eff(q)$\\
\verb$           +C3 * (1.0 + (-1.0)*m(j)*m(j)) * m(i)$
\end{minipage}

\bigskip
\subsection{Field dependencies}

\begin{figure}
\psfrag{m}{$\vec m$}
\psfrag{H_exch}{$\vec H_{\rm exch}$}
\psfrag{H_anis}{$\vec H_{\rm anis}$}
\psfrag{H_demag}{$\vec H_{\rm demag}$}
\psfrag{H_ext}{$\vec H_{\rm ext}$}
\psfrag{H_total}{$\vec H_{\rm eff}$}
\psfrag{rho_m}{$\;\rho_m$}
\psfrag{phi_m1}{$\Phi_{m,1}$}
\psfrag{phi_m1[boundary]}{$\Phi_{m,1,\rm boundary}$}
\psfrag{phi_m2[boundary]}{$\Phi_{m,2,\rm boundary}$}
\psfrag{phi_m2}{$\Phi_{m,2}$}
\psfrag{phi_m}{$\Phi_{m}$}
\psfrag{E_exch}{$E_{\rm exch}$}
\psfrag{E_anis}{$E_{\rm anis}$}
\psfrag{E_demag}{$E_{\rm demag}$}
\psfrag{E_ext}{$E_{\rm ext}$}
\psfrag{E_total}{$E_{\rm total}$}
\psfrag{dm/dt}{$\frac{d}{dt}\vec m$}

\includegraphics[width=0.8\textwidth]{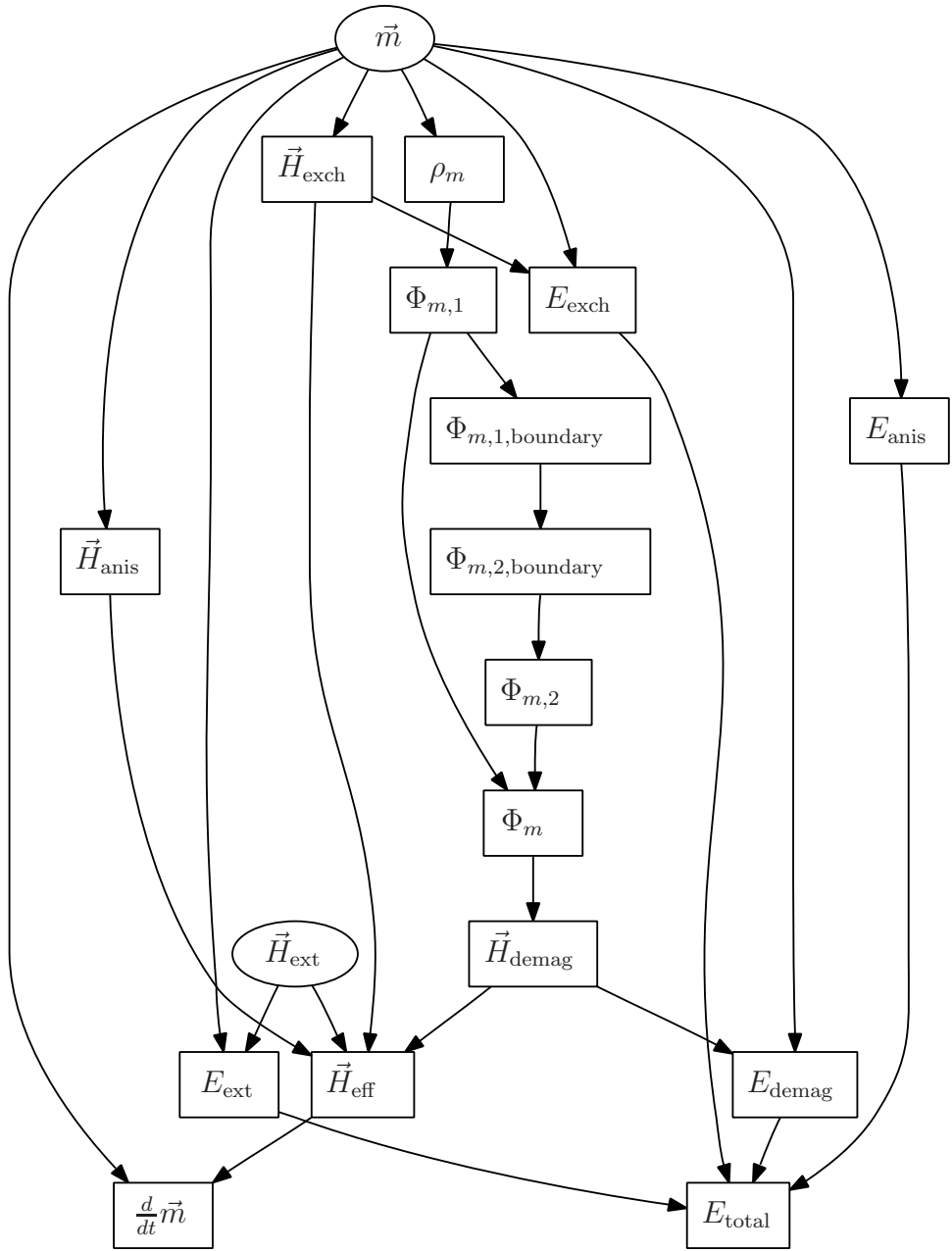}
\caption{Primary and auxiliary fields in micromagnetism 
(simplified).}\label{fig:tree}
\end{figure}

Auxiliary fields are computed -- sometimes via complicated dependency
hierarchies -- from the primary fields. Internal consistency in a
simulation framework demands that the manipulation of a primary field
followed by reading off the value of an auxiliary field which depends
on this primary field (directly or through intermediate quantities)
automatically triggers the recomputations necessary to provide the
updated value: the user of a simulation framework expects to see a
consistent picture where values of auxiliary quantities always are
automatically up to date. This is an important semantic aspect in
addition to the syntactic rules for defining systems. For
micromagnetism, the dependency tree of primary and auxiliary fields is
shown in figure~\ref{fig:tree} (ignoring dependencies between vectors
distributed across all computational nodes and sequential copies on
the master node).

\subsection{Other examples}

Due to the large number of different relevant physical effects -- some
of them computationally demanding -- micromagnetism is a particularly
heavyweight application of the \nsim{} compiler prototype. Simpler
physical models to which the \nsim{} framework should be even more
readily applicable include:

\begin{itemize}

\item Heat-flow and thermalization problems.

\item Combined electric/heat conduction problems.

\item Landau-Devonshire type models of ferroelectric materials.

\item Ginzburg-Landau theory of superconductors.

\item Turing's reaction-diffusion model of morphogenesis for pattern
  formation on animal skins.

\end{itemize}

A fully worked out example for the last item is given in
appendix~\ref{sec:appendix}.

\subsection{Physics beyond this language}

While the framework described here is sufficiently general to deal
with a quite broad class of problems in classical field theory, there
are a number of conceptual limitations. The presumably most relevant
ones are:

\begin{itemize}

\item It presently cannot handle problems that involve hyperbolic
  equations that require a Riemann solver, i.e. hydrodynamic problems.

\item The abstract framework does not yet provide approaches to couple
  field theory to non-field-theory physics (e.g. field-particle
  interactions as required to model particle detectors, or coupling to
  small-scale molecular dynamics models as used e.g. to study crack
  formation).

\item The framework also does not yet support specifications of
  dynamic (moving) boundaries.

\item Theories that require numerical models involving higher
  derivatives (such as the bi-harmonic equation in elastomechanics)
  cannot yet be expressed in the abstract operator language.

% This item is not really about physics:
%
%\item The compiler prototype presently is restricted to dealing with 
%finite element discretisations on simplicial meshes. Neither dynamic
%mesh refinement nor structured meshes, finite differences, or
%discretisation approaches such as the finite volume method are
%supported at present.

\end{itemize}

While these limitations can be overcome, they are beyond the scope of
this work.

\section{The {\nsim} compiler prototype}\label{sec:nsim-comp-prot}

Some of the technical aspects underlying the {\nsim} compiler --
especially those related to parallel computing -- warrant a more
detailed discussion. Again, we will frequently refer back to the
{\nmag} micromagnetic library whenever there is need an example to
explain some particular technique.

\subsection{Bookkeeping over field degrees of freedom}\label{sec:dofs}

A central data structure in the {\nsim} compiler is the {\em field
layout description}\footnote{For historical reasons and due to
{\nsim}'s FEM background, this is called a `{\em mesh with elements
(mwe)}' in the code.} The role of this field layout description is to
provide information on how to physically interpret the data contained
in a numerical vector that describes the state of a discretised field.

In the current implementation of {\nsim}, the field layout data
structure is constructed from a geometrical specification of the mesh
and additional information on the association of (abstractly
described) elements to mesh cells as a function of the cell's
associated region number. It would be rather straightforward to extend
{\nsim} in such a way that it also can handle other types of field
discretisations that are based on more specialized meshes (such as
regular grids), but this has not been implemented so far.

At present, the meshes are taken to be composed of simplicial cells of
arbitrary(!) dimension (i.e. tetrahedra in three dimensions,
pentatopes in four). An abstract element, which is associated to the
cells of the mesh by region, provides information about the degrees of
freedom belonging to this cell (and usually shared with neighboring
cells). An element may introduce multiple physical degrees of freedom,
and even multiple {\em types} of degrees of freedom at the same time
(e.g. three components of the electrical field strength, three
components of the magnetic field strength, and a mass and charge
density to describe the state of a plasma.) The actual location of a
degree of freedom may be given as a weighted average of the simplex
vertex coordinates (for higher order elements), and the corresponding
shape functions are given as polynomials in barycentric coordinates.

The individual physical degrees of freedom are taken to be real-valued
only. While it occasionally would be useful to support complex-valued
quantities as well, this runs into problems with {\nsim}, as the
underlying sparse matrix linear algebra library~(PETSc) cannot handle
both real and complex linear algebra at the same time within a single
program. This, however, is a purely technological rather than a
conceptual limitation. While a generalized multiphysics framework
certainly has to support scalar and vector degrees of freedom, it
should also provide higher order tensors of arbitrary rank to model
quantities such as stress, or flow of angular momentum (as
e.g. associated to spin-polarized currents). Furthermore, fields may
have internal degrees of freedom, i.e. some of the indices of a
multi-indexed quantity may not correspond to spatial directions. While
this situation is very common in non-abelian gauge theories (such as
quantum chromodynamics where QCD `photons' come in eight different
varieties), there are few (if any) well known examples in condensed
matter physics. This generalization nevertheless is so straightforward
that there would be no point in not supporting it as well, however.

The field layout data structure has the capability to provide
information about the type and indices of a degree of freedom as well
as its physical location in space (both in real coordinates as well as
in the form of a weighted average of the positions of mesh vertices),
its association to mesh cells and the corresponding shape functions,
the total volume associated to the `tent' spawned by the shape
functions, other degrees of freedom residing at the same site and
similar bookkeeping information.

The most important operations that use field layout information are:

\begin{itemize}

\item Extracting a particular subfield from the numerical data describing 
a multi-subfield field. This situation can arise when part of the full
configuration vector of all degrees of freedom (which e.g. is handled
by the time integrator) has to be split and passed on to individually
developed parts of the simulation that deal with specific physical
sectors (such as e.g. heat diffusion).

\item Re-Uniting subfields into a unified field vector (the opposite 
operation of the previous one).

\item Extracting some degrees of freedom of a given subfield into a 
shorter vector containing only a selection of sites (e.g. some
particular surface).

\item Inserting data from a restricted vector into a longer vector
(the opposite operation of the previous one).

\item Setting up bookkeeping information for calling machine code which 
was generated to handle non-linear purely local `reaction-type'
equations later on. (I.e. assembling tables of pointers into multiple
fields, one for every site, so that compiled code can access these
tables to find and identify the numerical values of individual degrees
of freedom living at the given site.)

\item Probing the value of a field at an arbitrary position, using the
analytic shape functions and interpolation.

\end{itemize}

Frequently, one wants to use field vectors which use different
names for individual subfields, but nevertheless are structurally
compatible, i.e. component-wise addition of data vectors is a sensible
operation. The most immediate example of such a situation is the
handling of an equation of the type:
\begin{equation}
\mbox{\tt M(j) = M(j) + dt * dM\_dt(j)}
\end{equation}

As was mentioned in section~\ref{sec:language}, usability experiments
have shown that a useful approach to handle this situation is to
implement {\em aliasing}: the {\nsim} user interface provides a
function (presently named {\tt mwe\_sibling()}) that is given a field
layout as well as subfield name substitutions and produces another
`sibling field layout' which is compatible with the original one (in
the above sense) but uses a different set of names for subfields. This
can be implemented very efficiently if care is taken that the degree
of freedom (dof) data structure does not contain a reference to the
subfield name but we instead use a separate function which maps a dof
plus a field layout data structure to a subfield name. Then, virtually
all of the entries of the field layout data structure (with the
exception of the vector containing the abstract elements associated to
mesh cells) can be shared between the original and the aliased field
layout, making aliasing very cheap in terms of memory requirements.
Furthermore, there are some slightly expensive entries in the field
layout data structure which are computed on demand when first needed,
such as information about all neighboring sites that carry a given
subfield.  (This is useful e.g. in micromagnetism to find out whether
the magnetization vector changes so much between adjacent sites that
simulation results cannot be trusted.) These can be shared in such a
way (by introducing an extra indirection level) that computing this
information for one field layout from a set related through aliasing
will also make this information available for all others at the same
time.

The present {\nmag} prototype discerns between primary and sibling
field layouts and does not allow secondary aliasing of aliased fields
in order to limit complexity. (This artificial restriction could be
removed.)

An important issue with fields is that a sparse matrix produced from
an operator via Galerkin discretisation does {\em not} represent the
linear mapping from discretized field vectors to discretized field
vectors corresponding to the original operator. This can easily be
seen by convincing oneself that the sparse matrix $I$ corresponding to
the identity operator is the matrix of integrals over products of
shape functions. In some situations, this is just what is required. In
others, we have to go from a vector of coefficients in dual space back
to a vector of coefficients of basis functions. In what follows, we
will refer to dual space data vectors as `discretised co-fields'. In
some situations, these play an important role as they encode
information about spatial integrals, e.g. when integrating total
outflux of a current across a surface. In others, we have to go from
co-fields to fields. While there is only one mapping that is
compatible with the natural mapping from fields to co-fields given by
the identity operator (i.e. solving the corresponding linear equation
system), this occasionally is not helpful, in particular as it
destroys locality properties: Considering the combined two-step
process of applying a matrix and solving an equation system, we
effectively work with a nonsparse linear mapping. If this contributes
to the Jacobian of the equations of motion, this matrix will
correspondingly become nonsparse and difficult to handle. For this and
other reasons, we usually go from co-fields to fields by dividing
every vector entry by the integral over the shape function associated
to the corresponding degree of freedom (cf.~\cite{Gardiner}).

During the development of the {\nsim} prototype, the strong
distinction between field and co-field data vectors that was
implemented by making use of the OCaml type system gradually turned
out to be more a hindrance than an advantage, suggesting that soft
flagging of numerical data as representing a field or a co-field may
be more appropriate.

\subsection{Dealing with field dependencies: the inference mechanism}

Auxiliary quantities are computed from the dynamical degrees of
freedom via either purely local (potentially nonlinear) operations, or
the application (or inverse application, such as in solving a Laplace
equation) of differential operators.

More complicated auxiliary fields that involve both differential and
non-linear operations usually can be regarded as being defined in
terms of a sequence of intermediate auxiliary quantities where each
step in the sequence only involves one of these two types of
operations. In effect, this means that there is a dependency tree for
auxiliary fields which is rooted in the primary fields. For the
micromagnetic model, this tree is shown in figure~\ref{fig:tree}.

Auxiliary fields have to be updated in two different situations:
either as part of the effort to compute the new velocity vector $\dot
y(t)$ from the configuration vector $y(t)$ when performing time
integration
% (or -- for these purposes -- equivalently, energy minimisation), 
or to reestablish consistency whenever the user changed the physical
configuration and reads out some auxiliary quantity from the
system. The control logic required to keep track of the necessary 
updates of derived entities mandated by changes to source entities
basically is the same that also underlies the `make' utility
program which deals with dependencies of source and derived files
belonging to a software project.

While {\nsim} provides such an inference engine which will keep track
of auxiliary fields that have become invalid, the present prototype
requires this dependency tree to be explicitly specified by the user.
As this information can be derived mechanically from the user-provided
definitions of the auxiliary quantities, it would be both feasible and
highly desirable to extend the compiler in such a way that the
dependency tree can be generated automatically in such a way.

In addition to the forementioned purely semantic aspects, the
dependency inference engine also handles buffer updates required to
transport numerical field data between individual processes in a
parallel computation: Multiple manipulations of primary fields will
not trigger parallel communication unless some parallel computation
is executed that uses these fields.

\subsection{Compiling differential operators}

In the {\nsim} prototype, the translation of a differential operator
to a sparse matrix is a three-stage process. In the first stage, the
operator string is parsed. (The corresponding parser being generated
from a grammar file by a conventional LALR(1) parser generator.) In
the second stage, the left hand side and right hand side field layouts
are analyzed for all relevant element combinations and a corresponding
internal table is constructed which contains functions that add
contributions to the sparse operator matrix if given the simplex to
act on. Both these steps are fast and performed only once. The
resulting table of functions is then used in a third stage to
initialize the sparse operator matrix, and can be re-used to
re-initialize the matrix multiple times (e.g. when a `middle field'
changes, or nodes are moved in space). In a parallel computation, the
internal function table is generated centrally and broadcast over all
nodes for subsequent distributed matrix initialization. This is
greatly facilitated by building the compiler on top of an
infrastructure that supports the serialization of functions (rather,
closures) into a network-transportable representation.

Internally, the various amendments that can be made to an operator
specification correspond to small modifications on the functions that
actually generate matrix entries. As the matrix populating function
resulting from stage~2 is fully polymorphic with respect to the few
matrix operations required (i.e. adding an entry, flushing, making a
matrix, finishing setup), higher-order programming trivializes most of
these modifications. E.g. periodic boundary conditions are implemented
by wrapping up the function making a matrix entry in such a way that
multiple entries are being made for degrees of freedom that have
multiple images on the mesh.

When the matrix-population functions are used to populate PETSc
matrices for the first time, some issues arise in conjunction with
pre-allocation techniques: The present version of PETSc does not
release memory that was pre-allocated when creating a sparse matrix,
and as {\nsim} at present does not pre-compute the number of entries
in each matrix row, so a trick is employed in order to release unused
memory which consists of copying the finished matrix and freeing the
original\footnote{The authors thank the PETSc developers for telling
them about this.}.

The compilation of differential operators is the most complex piece of
code within the \nsim{} codebase that lies at the core of the system,
with other parts having evolved around it.

\subsection{Solving operator equations}

Whenever a physical system involves very fast processes that come to
an equilibrium on time scales much shorter than those of interest, we
may benefit from employing potential theory to find those equilibria
that both depend on and influence the slower physics. In
micromagnetism, this certainly is the case for the electromagnetic
field: a change in magnetization at one end of a sample of size
$\ell=1\,\mu\rm m$ will take a characteristic time of about
$\ell/c\approx3\cdot10^{-15}\,\rm s$ to travel to the other end, while
characteristic time scales for magnetization dynamics are in the range
of $10^{-12}\,\rm s$. Hence, ignoring retardation and radiative
effects is a good approximation in the computation of the magnetic
self-interaction of the sample. The situation in micromagnetism is
considered to represent a generic case: the computation of some
auxiliary field (the demagnetization field strength $\vec H_{\rm
  demag}$) often involves a number of intermediate steps, some of them
requiring to solve a potential equation. In micromagnetism, the
problem of finding $\vec H_{\rm demag}$ corresponds to removing the
longitudinal modes of $\vec M$, as $\vec B=\vec H+4\pi\vec M$ and
$\vec\nabla\cdot\vec B=0$. This can be achieved either via Fourier
transformation on a regular lattice\footnote{The OOMMF~\cite{OOMMF}
  package is based on such an approach.}, or in the the
Finite-Element/Boundary-Element Method approach via a more involved
computation where two Laplace equations have to be solved for one
computation of $\vec H_{\rm demag}$: One to find the magnetic scalar
potential corresponding to $\vec\nabla\cdot\vec M$ with natural
boundary conditions, and one to extend the boundary corrections that
give `Dirichlet boundary conditions with zero potential at infinity'
from the surface to the interior.

While there are many problems where just a single Laplace equation has
to be solved (e.g. to get the electric current distribution in a
conductor), more involved situations similar to the one described
above are assumed to be somewhat common. At present, \nsim{} supports
these by allowing the user to define linear solvers on top of
operators and then specify (parallelizable) sequences of linear
algebra operations that may involve applying a solver, going from a
co-field to a field, etc.  While the underlying technology has been
implemented and can be employed to define even complicated
calculations, the problem of introducing a convincingly convenient
notation has not been addressed yet. The developers' present view is
that supporting a terse notation for two or three very common cases
that include obtaining a potential from a charge distribution as well
as removing some longitudinal part of a tensor field most likely
covers a large part of all relevant applications. The ability to
specify computation sequences in detail should be retained for unusual
situations.

\subsection{Nonsparse linear operators}

While every differential operator will produce a sparse matrix when
discretised in the straightforward way on a cell structure in space as
a consequence of its localization properties, there also are
situations when long-range effects which couple degrees of freedom at
arbitrary positions have to be modeled. Whenever this happens, the
system often becomes very challenging to simulate, doubly so in the
particular example of micromagnetism as in addition to long range
interactions, stiffness of the system becomes an issue as well.

From the physical perspective, long-range interactions are closely
related to very light bosonic exchange particles, with the most
prominent and therefore dominant case being that of a massless vector
boson such as the photon. Hence, those problems that require the
inclusion of long range interactions very frequently can be tackled by
employing potential theory in some clever way, with the standard
problem being the solution of Poisson's equation with Dirichlet
boundary conditions at infinity. This can be regarded as a third class
of boundary conditions of great practical importance in addition to
the more straightforwardly defined Dirichlet and von Neumann boundary
conditions. On unstructured meshes, a widely used technique to solve
this class of problems was given by Fredkin and Koehler
in~\cite{Fredkin1990}. Essentially, their hybrid Finite
Element/Boundary Element method boils down to the introduction of a
square $N\times N$ matrix, $N$ being the number of boundary degrees of
freedom, that encodes all the information about boundary-boundary
interactions. For potential problems using first order elements, an
analytical formula given by Lindholm~\cite{Lindholm1984} that
describes the potential of a triangular dipole layer with linearly
varying source density can be used to compute the matrix elements.
While this technique works sufficiently well to be employed e.g. in
micromagnetism, there are a number of problems:

\begin{itemize}

\item Even for moderately sized simulations, the boundary element matrix 
may require a considerable amount of memory, as well as setup time.
($10\,000$ surface nodes would require $\sim 800\,{\rm MB}$ of RAM
using standard 64-bit floatingpoint element representations). This is
especially a problem with thin film geometries.

\item Lindholm's formula suffers from numerical cancellation problems in 
some geometric configurations.

\end{itemize}

In a FE/BE implementation, the latter problem may be alleviated to
some degree by internally using extended floatingpoint numbers for the
computation, but this often is impractical. A numerically more useful
equivalent of the original formula hence would be desirable.
Concerning the former issue, there are approximation techniques to
reduce memory footprint such as using hierarchical matrices which are
based on a clustered approximation of propagators $G(x-y)\sim\sum_k
c_k G(x)\cdot G(y)$~\cite{Hackbusch2003}. The {\nsim} prototype supports
explicit computation of a nonsparse $N\times N$ boundary element
matrix for first order elements based on Lindholm's formula and also
can utilize the HLib~\cite{HLib} library to use matrix
compression\footnote{Due to license issues, HLib does not come with
  {\nsim}, but has to be obtained and installed separately.}.  Support
for quasi-long range interactions, i.e. forces mediated by particles
whose mass corresponds to a length scale somewhere between the cell
and the mesh size is not implemented at present.

\subsection{Periodic boundary conditions}

The term `periodic boundary conditions' is used for two conceptually
different ideas: on the one hand to reduce the finite-size artefacts
in computer simulations of the statistical properties of a sample, and
on the other hand to simulate some of the relevant processes
associated to a single cell in a repetitive sample. Subtle issues
arise when systems made up of very many periodic copies of a
fundamental unit are to be simulated employing periodic boundary
conditions while they also feature long range interactions. In such a
situation, it usually becomes necessary to amend the numerical model
with relations between the wave-vector-zero components of
fields. Again, this situation is easy to understand through the
example of micromagnetism: While it may be possible to learn a lot
about the behaviour of a system made up of $1\,000\,000$ identical
cells by simulating a single cell in the interior with periodic
boundary conditions, the relation between the average of the
magnetization $\langle\vec M\rangle$ and the average of the
demagnetizing field $\langle\vec H_{\rm demag}\rangle$ over the cell
are still determined by the shape of the (faraway) boundaries of the
sample only. There is some initial support for these `macro-geometry'
effects in \nsim{}, as they are relevant for 
micromagnetism~\cite{Macrogeometry}, but it is not yet entirely
clear whether this is sufficiently generic for other applications
where such situations arise.

Subtle long-range issues aside such as those described above, periodic
boundary conditions are supported by allowing meshes to be associated
with information on the identification of vertices, from which
identifications of sites also follow (which is especially relevant for
higher-order elements). A generalization that would often be useful
but is not yet supported would be to allow arbitrary affine-linear
transformations between the fields living at different periodic copies
of a site. This would e.g. allow to simulate systems with axial
symmetry by imposing suitable `periodic boundary conditions' on a
wedge, or to set up systems in which potentials on opposite faces are
identified up to a constant, etc. This approach would, however, raise
a number of somewhat tricky technical issues related to the automatic
generation of Jacobi matrices.

\subsection{Local non-linear equations}

The other elementary building block of field operations (apart from
linear, differential field operators) is arbitrary (i.e. usually
non-linear) {\em site-local} manipulations of fields. In combination,
and through the introduction of intermediate auxiliary fields (such as
e.g. $H_{\rm eff}$ in micromagnetism), these basic operations cover
the vast majority of field operations encountered in applications.

The notion of {\em site-local} operations encompasses all
manipulations on a set of fields which are to be performed at every
mesh site in the same way, and operate on the subfield components of
these fields which are located at the specific site only. This
basically corresponds to `considering position dependency as
implicitly being understood' in physical notation when giving an
equation such as the Landau-Lifshitz and Gilbert equation in the
form~(\ref{LLGtensor}).

We may want to use site-local manipulations that involve highly
unusual operations which cannot be expressed easily in arithmetic
notation. (For example, we may want to use different manipulations
depending on whether some temperature is above or below a critical
value.) As a consequence, maximum flexibility demands the capability
to allow arbitrary user-specified code to specify local physics. This
code nevertheless has to be executed fast, therefore the {\nsim}
prototype compiler resorts to employing runtime compilation
techniques, as explained in section~\ref{sec:fastfields}.

Before sending user-supplied code to the C compiler, {\nsim} wraps it
up in a block if {\tt \#define}/{\tt \#undef} macros which allow the
user code to refer to local subfield components directly by their name
and indices, regardless of the field data vector they are located
in\footnote{The observed utility of this approach was the underlying
  reason for implementing `field aliasing', cf.
  section~\ref{sec:dofs}}. Also, {\nsim} provides C macros with which
presence of a particular subfield at a given site may be tested, as
this may depend on the site in heterostructures.

From the perspective of user-supplied local code, the {\nsim}
framework provides a context in which to execute this code over all
sites of a mesh on a number of fields (and co-fields) via wrapping
definitions, as well as behind-the-scenes bookkeeping over
site-dependent offsets into the field data vectors, plus automatic
compilation-on-demand.

While it is crucial to provide maximal flexibility to the user through
the option to specify arbitrary code for site-wise execution, it is
just as important to also provide an alternative interface which,
rather than asking the user to write C code, allows specification of
equations in algebraic notation. This is achieved by employing a
LALR(1) parser generated by OCaml's equivalent of `yacc' from a formal
grammar specification and translating algebraic notation to C code
internally, which then is provided to the general mechanism. Also,
algebraic notation allows auto-generation of derivatives of equations
of motion (through some internal symbolic algebra) as needed for the
Jacobian which should be known (at least approximatively) to
successfully employ some time integrators, such as
CVODE. (Cf. sections~\ref{sec:jacobi} and~\ref{sec:sundials}.)

\subsection{Computation of the Jacobian $d\dot y_j(t)/dy_k$}\label{sec:jacobi}

A not too uncommon situation is that the physical system under
investigation may be numerically challenging due to its stiffness. In
micromagnetism, this happens due to a gap of more than two orders of
magnitude between the characteristic time scales of the reaction of
the system to a strongly localized perturbation. This comes from
different physical effects (in the simplest case, the long-range
demagnetization field and the extremely strong effective exchange
coupling that tries to make closely spaced magnetic degrees of freedom
align and move in unison). There are a number of code bases that
implement time integration algorithms for stiff systems which often
can be used in a `black box' fashion. Normally, the user of such a
time integration library has to provide both a function that computes
the velocities, given some particular state of the system, as well as
a function that computes the Jacobian. Roughly speaking, the Jacobian
helps to find a transformation from the fundamental degrees of freedom
to a more useful basis which allows to separate fast processes (`high
frequency ringing' of the system) which often are not excited in
physical systems from the more interesting collective slow dynamics.

The Jacobian usually is a function of the configuration and time, and
as it is expected to be costly to compute, the time integration
library will try to avoid recomputation of the Jacobi matrix as much
as possible. (Evidently, some heuristics has to be used here, as it
cannot be known beforehand how many processor cycles the computation
of the Jacobi matrix will take in a particular situation.)

In principle, the Jacobi matrix can be computed automatically when
given the equations of motion. In reality, most physical simulations
use hand-written low-level code (usually C or FORTRAN) to implement
the computation of the numerical matrix entries. For systems with more
complicated equations of motion (such as micromagnetism), the
implementation of the code that computes the Jacobian can become a
rather tedious and frustrating process that is prone to errors which
here unfortunately are easy to make but fiendishly difficult to track
down.

For this particular reason, {\nsim} provides capabilities to
automatically generate the Jacobian from the symbolically specified
equations of motion. While it would in principle be possible to just
use these as they are and symbolically take second derivatives, one
discovers in practice that it makes sense to introduce some additional
flexibility in the specification of the Jacobian. One important issue
is that one may use equations of motion for the Jacobian (which only
needs to be approximatively right) which are slightly different from
the system's actual equations of motion. For example, if extra
nonphysical terms have been introduced for the sole purpose of
numerical stabilization, one will usually want to simplify the
Jacobian by not including these.  A second type of important
simplification that must be supported is to allow the use of
approximations to derivatives of dependent fields. In the particular
case of micromagnetism, it is useful to approximate $\delta H_{\rm
eff}/\delta M$ by $\delta H_{\rm exch}/\delta M+\delta H_{\rm
anis}/\delta M$, omitting the contribution from the demagnetizing
field (which would at the same time be very costly to compute, not
expressible through a sparse matrix contribution to the Jacobian, and
not a source of fast dynamics).

The present internal design of the component providing symbolic
auto-generation of the Jacobian in the {\nsim} prototype presumably
has not settled down to its final form yet, as it still appears to be
somewhat unwieldy in some situations. At present, the interface works
in such a way that the user specifies both an equation of motion, such
as e.g. a string representation of the Landau Lifshitz and Gilbert
equation, as well as additional information on contributions to the
derivatives (with respect to the configuration degrees of freedom) of
the fields that appear in the equation of motion. This is given as a
list, one entry for each field that enters the equations of motion
(with meaningless entries such as $\delta \vec M/\delta \vec M$
ignored).  Each list entry itself is a list of contributions which
will be summed and either provide a local non-linear (at present, only
polynomial) equation, or a name of a sparse operator. The former case
is used for providing the contribution to $\delta H_{\rm eff}/\delta
M$ that comes from the anisotropy field strength $\delta H_{\rm
  anis}/\delta M$, while the latter is used to for providing the
$\delta H_{\rm exch}/\delta M$ contribution.

The system then internally `precompiles' this information to a
function that can be called in order to (re-)populate the Jacobi
matrix. In a parallel computation, this is handled as with other
sparse matrices: the population function is generated centrally and
then distributed over the cluster. Evidently, such a matrix population
operation also requires making the data necessary to work out the
individual contributions to all nodes (i.e. a MPI {\tt allgather()}
operation on relevant fields). Some complications arise as
recomputation of the Jacobian is triggered by the time integrator: on
the one hand, rather than just updating the Jacobian $J$, the sparse
matrix $\mathbb{1}-\gamma J$ (with $\gamma\in[0,1)$ given) has to be created
and updated (in parallel) as well. Furthermore, when doing simulations
with periodic boundary conditions, the situation is complicated by the
necessity not to duplicate entries in the state vector of the time
integration library used by {\nsim} on the one hand and the desire to
directly use sparse matrices which use duplicate entries on the other
hand. Hence, the function that is triggered internally to recompute
the Jacobian has to do some automatic index mapping to hide this mismatch.

An earlier approach used in {\nsim} to compute the Jacobi matrix~$J$
involved the pre-computation and distribution of a data structure that
provided a symbolic polynomial for the computation of each nonzero
matrix entry. This was soon abolished as it turned out to use
prohibitive amounts of memory.

\subsection{The `physics engine' abstraction}

When setting up a simulation, \nsim{} translates a complete
specification of all field layouts, fields, sparse and nonsparse
operators and linear solvers, as well as local equations and field
dependencies to an object-like data structure that provides methods to
set and read off fields, and to advance time\footnote{In the present
version of the \nsim{} prototype, this is (for historic reasons) named
a `linear algebra machine'}. While this `monolithic' approach greatly
simplifies the compiler, it ultimately should be broken up so that it
becomes possible to freely add (and remove) operators to a simulation
object after initial setup. Technically, this would be possible with
rather little effort in the present {\nsim} codebase, as most of the
internal bookkeeping of dependencies of data structures on one another
are compatible with such a restructuring.

\subsection{Parallelized time integration}\label{sec:sundials}

While the idea may be seductive to simplify the code structure of a
parallel multiphysics simulation framework by employing parallel
computing only for computationally expensive linear algebra
(e.g. solving linear equations) and keeping time integration
centralized in the controlling process, one finds that for challenging
problems that involve stiff dynamics (such as micromagnetism), such an
architectural design becomes a major bottleneck: Measurements show
that about as much numerical effort is required to deal with stiffness
in the time integrator as is needed to compute all the expensive
auxiliary fields. Hence, time integration has to be fully parallelized
as well.

In some situations, time integration has to respect additional
constraints which must not be violated by numerical drift. Such a
situation occurs in {\nmag} with the magnitude of magnetization and is
dealt with by introducing an additional term in the equations of
motion for numerical stabilization, as explained in
section~\ref{sec:micromagnetism}.

The \nsim{} prototype presently utilizes the readily parallelizable
CVODE time integrator from the Sundials~\cite{Hindmarsh2005} library.
As this is an adaptive algorithm whose implementation in the Sundials
library experiences problems with fixing step sizes, the \nmag{}
application also contains a fixed-step Heun integrator for
applications where this is required. The incentive to include this
alternative comes from the desire to simulate thermal fluctuations
based on Langevin dynamics. However, as the duplication of degrees of
freedom due to periodic boundary conditions needs special handling in
time integration, this alternative integrator does not yet support
periodic boundary conditions.

\subsection{User interaction}

The idea of using computer simulations to study some particular aspect
of the behaviour of a physical system intrinsically calls for a
certain level of flexibility. One rather obvious reason is that one
would like to be able to automatize the processing of simulation
results to tables, diagrams, and other images. In addition to this,
there are situations where one would like to use a bit of automatized
choreography for performing a parametrized sequence of simulation runs
(`computational steering'). A prime example for this would be to run a
series of individual micromagnetic simulations where the externally
applied magnetic field strength varies in a controlled way in order to
produce some particular switching sequence. Here, one would ideally
want to be able to wrap some control and post-processing code around
the simulation core in such a way that the mapping of a given
switching sequence to a number of diagrams which describe the physics
of the process is fully automatized.

\subsubsection{Standalone programs vs. libraries}

While such situations that basically ask for a Turing-complete level
of flexibility are rather widespread, many specialized software
packages have been written where the a design mistake has been made to
initially underestimate the amount of flexibility that eventually
would be required and trying to mend this progressively in incremental
steps.  Typically, such systems gradually evolve from being able to
read very primitively structured configuration files towards using a
LALR(1) parser in order to deal with more complex configuration files
towards supporting some rudimentary form of variables, arithmetics,
and control structure. Subsequently, a need arises for container data
types such as arrays (which also introduces a number of tricky memory
management issues), then file I/O, and finally the capability to
interface C libraries and a debugger.\footnote{This phenomenon, which
  has been encountered with word processors, CAD software, ray
  tracers, web browsers, databases, is commonly referred to as {\em
    Greenspun's Tenth Rule of Programming}: `Any sufficiently
  complicated C or Fortran program contains an ad hoc,
  informally-specified, bug-ridden, slow implementation of half of
  Common Lisp.'}

A reasonable approach to avoid this problem is to provide maximum
flexibility from the beginning through designing the system in such a
way that it provides all of its special capabilities in the form of a
software library and interface this to some existing popular
programming language with a large user base. 

\subsubsection{The Python scripting language as an especially attractive option}

For {\nsim}, the design decision to utilize Guido van Rossum's
`Python' scripting language~\cite{Python2003a} (and hence give {\nsim}
the form of a Python library) was made based on a number of
circumstances:

\begin{itemize}

\item {\em Accessibility}: A file that exclusively consists of a
  sequence of lines of the form ``{\tt parameter = value}'', possibly
  interspersed with comments, already is a valid Python script.
  Therefore, simulation specifications can be as simple or complicated
  as one likes. It is easily possible to let a proficient Python
  programmer add a very small (``level 4'') layer of Python code on
  top of the {\nmag} system to turn it into a very specialized tool
  for magnetic simulations that also is extremely simple to use by
  less computer literate users -- e.g. for automatically producing a
  large number of graphs of hysteresis loops.

\item {\em Availability}: Python is widely available, under a free license, 
on a large number of platforms.

\item {\em Familiarity}: Python (unlike Lisp/Scheme) has a simple syntax 
that is easily picked up by anyone who has had previous exposure to virtually
any other imperative programming language (such as C, Perl, Matlab, or even
the shell).

\item {\em Popularity}: Python both has a sizable user base, especially 
in engineering, and is rather accessible for both the inexperienced casual
user and the expert.

\item {\em Power}: Python supports enough abstract concepts
(e.g. container data structures such as dictionaries) to make the 
implementation of more sophisticated algorithms for control and data
processing a realistic option.

\item {\em Engineering Relevance}: There are a number of Python
libraries that make it a viable substitute for Matlab.

\end{itemize}

Python support also brings some additional secondary benefits:

\begin{itemize}

\item {\em Interactivity}: It is possible to use the Python interface 
in an interactive execute-commands-as-they-are-given fashion to explore
and hence get a more intimate understanding of some particular system. 
{\em With \nsim{}, this also works for parallel computations: interactivity 
and parallelism do not exclude one another}~\cite{fischbacher:07D527}.

\item {\em Libraries}: There is a large number of third-party Python 
libraries which easily can be used to provide additional capabilities
such as database support, image generation, automatic website
generation, and much more.

\item {\em Debugging}: Experienced Python programmers can 
utilize advanced debugging capabilities which usually are hard to find
in application-specific languages. With Python, it even becomes
possible to interactively inspect the situation in the context where a
problem occurred by escaping to a command line from deep within the
program.

\end{itemize}

However, should Python go out of fashion at some point in the future
(as so many other scripting languages have done), {\nsim} is designed
in such a way that it would be possible to include support for another
user interface level language with reasonable implementation effort.

\subsubsection{Python libraries}

While from the perspective of a Python programmer, \nsim{} behaves
like a Python library, it usually is executed as a special program
that behaves like an extended Python interpreter, rather than just
starting the standard Python interpreter and then importing the {\tt
nsim} module. There are a number of reasons related to the behaviour
of different subsystems utilized by \nsim{} that suggest such a
design. In particular, some implementations of the MPI framework use
extra command line arguments to communicate parallel execution
parameters which may clash with parameters used by the Python
interpreter\footnote{This also is the reason for the existence of the
{\tt mpipython} program}. As this may cause some inconvenience to
Python programmers, and as these issues are merely technical
subtleties rather than truly fundamental limitations that would
prevent \nsim{} from being turned into a proper Python extension
module, this is actively being worked on at the time of this writing.

\subsubsection{Dynamic fast code and scripting languages}\label{sec:fastfields}

One general problem inherent in a scripting language approach is
that situations may arise where small pieces of user-specified code,
such as e.g. a function describing a scalar potential, or a non-linear
local field equation, has to be evaluated so often that only a
compiled machine-code version would be sufficiently efficient not to
dramatically slow down the simulation. On the other hand, it is highly
desirable to retain full programming flexibility rather than
artificially restricting the types of functions that can be specified.

There are a number of high level languages which both can be used
interactively and produce fast machine code at the same
time.\footnote{Two widely known examples would be the CMU Common Lisp
  compiler~\cite{cmucl} and the SML/NJ system~\cite{smlnj}} However,
producing machine code from scripting language code is not a realistic
option as scripting language popularity seems to be strongly
influenced by fashion trends: one would have to tie the system so
strongly to one particular user interface language that switching over
to a different one becomes unrealistic. Also, in the specific case of
Python, the Python compiler, Psyco~\cite{Rigo2004}, does not (at the
time of this writing) support code generation for machine
architectures other than x86-32, i.e. cannot generate fast 64-bit
code.

Therefore, {\nsim} pursues a different approach: virtually all
situations where user-specified pieces of code in a field theory
simulation system have to be fast can be regarded as special cases of
routines that implement a $\mathbb{R}^n\rightarrow \mathbb{R}^m$
function, potentially with additional internal state, potentially
utilizing an extra level of pointer indirection to get at numerical
data. So, the {\nsim} simulation compiler provides capabilities to
regard a user-specifiable string as a piece of C code and turn this
into a callable function. Internally, this works by dynamically
generating a C source file which is sent to the C compiler to produce
a shared object which then is linked lazily -- i.e. when such a
function is used for the first time -- (using the {\tt dlopen()} and
{\tt dlsym()} functions from {\tt libdl}) into the running simulation.

Site-wise execution of C code generated from a specification of a
local polynomial in fields (such as anisotropy energy in
micromagnetism) uses this mechanism in conjunction with a number of
autogenerated C preprocessor macros as well as an extra array of
pointers to floatingpoint data which provide access to the
site-relevant field entries.

This mechanism -- implemented in the {\tt fastfields} module in the
{\nsim} code -- satisfies some evidently very desirable properties:

\begin{itemize}

\item The C compiler is not called unnecessarily: When 100 functions that
have to be compiled are defined one after another before any one of them
is called, only a single call to the C compiler is being made. (This
is achieved by lazily deferring compilation to the latest possible
point in time.)

\item Whenever recompilation is initiated, functions that have become 
inaccessible through forgetting all references to them (i.e. they have
become garbage) automatically disappear from the dynamically generated
library. (So, re-defining such a function 1000 times during the course
of a program will not increase memory usage.)

\item At any given time, only a single dynamically compiled shared
object module will be linked in. (This means that defining and calling
new functions (which induces recompilation) will have to unlink and
reset the symbol-to-address associations of other dynamically compiled
functions (which will once again have to be looked up using {\tt
dlsym()} after relinking).)

\end{itemize}

The major drawback of the present implementation is that a single
syntactic error in one user-specified C function will break all such
functions (as it breaks recompilation). The most promising approach
towards this problem presumably is to add an extra checking phase
where the compiler is used on every single function as it is specified
for the first time to find code errors and reject it (i.e. compile it
to a function that raises an exception) if this situation occurs.

\subsection{Parallelization}

The capability to use parallel computation both to speed up linear
algebra as well as non-linear (site-local) operations as well as to
distribute the memory requirements of data structures should be
considered more a necessity than merely a desirable feature. The
underlying reason is that parallel computation is widely used in
simulations these days, hence a nonparallel simulation framework would
suffer from severely restricted applicability in view of the present
range of problems studied by computer simulations. This is much more
so due to the increasing availability of multicore CPUs.

\subsubsection{The MPI execution model}

Concerning parallelized number-crunching (as opposed to e.g. massive
parallel lookup and search operations over clusters of potentially
unreliable machines, i.e. search engine technology), the situation
that emerged over the last years is strongly dominated by the
availability of a flexible unified interface for efficient data
exchange between processes partaking in a parallel computation, namely
the MPI libraries. Very broadly speaking, MPI-based programs consist
of a mixture of `ordinary' commands as well as `collective'
commands. The role of the latter can be regarded as providing a
refinement of the execution semantics of the underlying programming
language (most often, C) by defining a kind of `must happen in unison
across all processes participating in the computation' generalized
execution flow concept. Traditionally, this is usually achieved by
writing MPI-based programs in such a way that all machines run the
same program code which executes the same sequence of statements on
all cluster nodes, very slightly seasoned only with node-number
dependent range selections that ensure every process works on its
share of the (often, linear algebra based) distributed workload. In 
particular, the entire parallel communication choreography is 
pre-determined right from the start of the program.

We will call this execution model the `choir model': with the image in
mind of songtext distributed at the beginning to all the singers which
precisely defines when to sing which tune, where to loop (and how
often), etc.

The obvious advantage of this model is that it is very easy to
comprehend, implement, and debug. In principle, more general
distributed algorithms could be (and often are!) far more
complicated. For evident reasons, the choir model is the dominant
model for parallelized number crunching.

Unfortunately, this kind of execution semantics is fundamentally
incompatible with the idea of providing maximum user flexibility. In
particular, as soon as there is any way how program flow could depend
on nonpredictable external input (such as, for example, provided
interactively by a human user), the choir model no longer can be
applied. Instead, we must resort to a more sophisticated model where
not only the data to be worked on, but also information about the
nature and order of operations on that data has to be distributed at
run time across all the computation nodes.

Furthermore, subtle issues arise in conjunction with dynamical
distributed resource management. While memory management for
conventional parallel number crunching programs usually is rather
trivial, allocating all relevant data structures in the beginning,
populating matrices, doing the computation, freeing all resources in
the end, the situation is very different for highly dynamical
applications. When does a vector which is distributed across all nodes
of a cluster become inaccessible, and if this happens, how do we
ensure this resource is freed in unison across all the cluster nodes?

The design decision underlying the {\nsim} compiler prototype is that
the user should not at all have to worry about any issues related to
parallel execution. With the present implementation, this goal has
been achieved to the largest possible extent. The only parts of the
system that are not fully parallelized are:

\begin{itemize}
\item Loading the mesh.
\item Working out the partitioning of mesh vertices across cluster nodes.
\item Some parts of the setup phase, most importantly working
      out the offset tables for local site-wise C code execution.
\item Writing simulation data.
\item Computations involving the compressed boundary element matrix
      that are based on HLib.~\footnote{At the time of this writing,
        parallelizing HLib is actively being worked on.}
\end{itemize}

All these cases raise somewhat tricky conceptual issues on the
expected behaviour of a parallelized program that have to be resolved.

\subsubsection{Mesh partitioning}

The \nsim{} compiler utilizes the Metis/Parmetis~\cite{ParMETIS}
library to both work out node re-orderings that improve the
performance of sparse matrix operators as well as node partitionings
across the cluster. Inter-process communication is used to distribute
both complete meshes as well as associated partitioning information
across all cluster nodes.

\subsubsection{Shielding the user from MPI}

The \nsim{} prototype enables the user to set up simulations that
allow parallel execution without ever having to worry about any
details associated with multi-machine execution flows. One particular
issue is that of allowing users to debug their own code: if the
complexity associated with parallel execution and collective MPI
commands showed up at the user level, this would seriously interfere
with the user's capability to inspect and analyze some particular
situation interactively in a debugging session. Rather, the user
should never have to concern himself with questions such as which
statements he can execute interactively and which are forbidden as
they would require coordinated execution of MPI-collective commands:
things should just work as one would expect with the model of ordinary
single-process execution flow in mind.

This is indeed achievable, but requires extensions to distribute not
only numerical data across the cluster, but information about execution
sequences as well. The basic idea is to let the master node execute
the simulation script and make all the other nodes go into `slave
mode' where they receive and subsequently execute data, command
sequences (akin to OpenGL display lists), and commands from the
master. The commands understood by the slaves include:

\begin{itemize}

\item Shutdown of the system and subsequent termination of the program.

\item Parallel linear algebra resource allocation
 (for vectors, matrices, linear solvers, etc., but also for meshes,
 field layouts and functions that populate sparse matrices).

\item Command sequence registration: a sequence of
operations which manipulate distributed resources is registered
under some particular name for later execution (so that later on,
a command sequence can be executed across the cluster without
having to employ parallel communication to send out every single
command as it is executed).

\item Execution of a previously registered command sequence.

\item Parallel (linear algebra and command sequence) resource de-allocation.

\end{itemize}

It is important to note that, while the slave processes contain all
the machine code of the core~{\nsim} executable, they do not run the
user's program but are driven exclusively by commands sent via MPI
from the master node.

One could call this the `organ model' (in contrast to the `choir
model') of parallel execution: the computation utilizes hardware that
is spread out over a large area, and the musician has full freedom to
interweave at will and from a central console both programming (using
organ stops) and playing.

The implementation of the {\nsim} core internally is based on the
Objective Caml system, which provides a proper substitution-aware
function concept that is much more powerful than the more simplistic
concept of a `routine' (which is employed by most compiler languages,
such as C, and sometimes confusingly called a `function' there as
well). In particular, OCaml supports (as an experimental but highly
useful feature) the serialization of functions to strings in such a
way that only the implicitly fixed contextual (`closure') parameters,
but no compiled machine code, have to be sent over the network. This
way, a function which was defined through parameter-fixing of a more
generic function on the master node can easily be made available on
the slave nodes as well.  This is used e.g.  to gain maximum
flexibility in distributing functions over the cluster that populate
sparse matrices. However, one consequence is that {\nsim} in its
present form can not work on heterogeneous multi-architecture
clusters, as one should not expect Objective Caml data structure
serialization to be able to cross architecture boundaries.

A {\nsim} script can be either started in single-process mode
\[
\mbox{\tt \$ nsim simulation.py}
\]
or under MPI control (here, on four nodes) to execute in parallel:
\[
\mbox{\tt \$ mpirun -np 4 nsim simulation.py}
\]

Unfortunately, some MPI implementations are somewhat idiosyncratic in
the way they provide parallelism to C programs: what effectively
happens is that another pre-startup layer gets wrapped around the
program's {\tt main()} function that parses and interprets MPI-related
command arguments. This is then used in conjunction with a utility for
remote login (such as {\tt ssh}) for automated coordinated startup of
individual processes. This has the unfortunate effect that it can
interfere with argument parsing of scripting language interpreters,
such as Python.

It is important to note that the startup code of the standalone
{\nsim} executable has to discern between a number of different
situations: using {\nsim} to run a user simulation program whose
filename is given as a command argument induces a different startup
mode than starting {\nsim} interactively. Furthermore, both
interactive and script mode may be used when running {\nsim} in a
distributed fashion under MPI control. (Interactive MPI-based parallel
execution is a highly uncommon feature, but possible with {\nsim}.)
Note that in parallel computations, regardless of whether script or
interactive mode is invoked, slave processes will always go through
special startup code to enter slave mode.

\subsubsection{Centrally coordinated parallel linear algebra}

The `organ' execution model described in the previous section requires
the distribution and management of both linear algebra resources
(vectors, matrices, linear solvers) and control resources (execution
sequences of linear algebra operations). For the {\nsim} prototype,
the MPI-based PETSc library was chosen to provide parallel linear
algebra capabilities. In MPI parlance, PETSc creates an own
communicator data structure to separate its own parallel communication
from other MPI-based communication the program may also have to
perform. The {\nsim} core uses an additional communicator for its own
coordination and management-related messages.

The most important issue with dynamic parallel execution is dynamic
resource management: both Python and the Objective Caml system provide
automatic resource de-allocation capabilities (of different
sophistication) that have to be extended to parallel
resources. Fortunately, it is not necessary to address the interplay
of Python's dynamical memory management with parallel resources, as
all parallel resources can be managed internally by the OCaml system
and proper behaviour with respect to Python code is then induced
automatically through the Python-OCaml interface.

The basic idea underlying dynamical resource management is that, as
the master node has the sole responsibility for execution flow,
garbage collection on the master node will be the sole authority on
dynamic de-allocation of distributed resources. This is effected
through addressing all distributed resources in an uniform way through
{\em dynamic resource handles} which basically represent names of
dynamical resources to which a finalizing function is associated -- on
the master machine only -- that will be executed upon garbage
collection. Every process participating in a parallel computation
(including the master) maintains a dictionary that maps dynamic
resource handles to the actual resources (i.e. this machine's share of
a distributed vector, matrix, or a distributed script, etc.) When one
dynamical resource (i.e. a command sequence) refers to another one
(i.e. a vector), it does so by using this dynamical resource handle,
so garbage collection of the command sequence resource can induce the
de-allocation of other resources referred through it.

On the slave nodes, dynamic resource handles are nothing but just 
keys into a dictionary containing the corresponding resources, without
finalizers attached. They are being managed (i.e. allocated and
de-allocated) exclusively according to resource management commands
received from the master. Here it is important to know that the OCaml
data serializer works in such a way that turning an entity such as a
distributed resource handle into a string and sending it over the
network will strip it from its finalizers. Hence, parallel command
sequences can be serialized and distributed from the master to the
slave processes conveniently without having to worry about spurious
finalizers being introduced on the slaves.

On the master, de-allocation of a distributed resource has to induce
sending out the corresponding de-allocation commands to the slave
nodes, as this may involve the execution of collective MPI
commands. As garbage collection can and will happen at arbitrary
points throughout the execution of a program, for example in the
middle of a registered sequence of linear algebra commands which
involves MPI-collective operations before and after the garbage
collection, the master {\em must not} induce distributed de-allocation
from within a distributed resource handle finalizer (as this may upset
the synchronity of the sequence of MPI-collective commands across
machines), but instead defer the actual sending of de-allocation
requests to the next point in time when slaves are again ready to
receive commands from the master. The obvious drawback of this
approach is that there are situations where distributed resources
refer to other distributed resources in such a way that de-allocation
cannot happen simultaneously but will artificially be spread out over
multiple cycles of garbage collections and entries to the parallel
command distribution code. For real world applications, however, this
does not pose a problem.

Evidently, the resource tables that map distributed resource handles
to the actual resources must not use those actual distributed resource
handles as keys for which finalizers were registered (otherwise their
presence in these tables would prevent them from ever being
collected!) While one may employ weak reference tables to circumvent
this problem, a simpler approach is to use copies of the distributed
resource handles as table keys which do not have finalizers
registered.
\subsection{On the implementation of {\nsim}'s core capabilities}

Underneath the user interface, {\nsim} has to provide not only
numerical capabilities, but also some rather sophisticated bookkeeping
as well as symbolic transformations (e.g. analytically working out
integrals over simplices). Evidently, it is highly desirable to have
these computationally demanding parts implemented in a fast compiler
language. The major criteria that influence the choice of a viable
compiler to base such a system (developed in an academic environment)
on are:

\begin{itemize}

\item The compiler should be sufficiently well supported to get
  feedback from the developer team when low level technical issues
  (such as interaction with MPI) arise.

\item The compiler should be able to produce efficient code for the
  most widespread platforms.

\item The language should provide basic capabilities that are highly
  useful when implementing symbolic algorithms, including lambda
  abstraction, closures, tail recursion, and a proper garbage
  collector.

\item The language should be sufficiently easy to pick up for new PhD
  students on the project (which may be physicists or engineers rather
  than computer scientists), so that they can start to be productive
  very early on.

\item The language should allow interfacing specialized C libraries
  (e.g. for parallel sparse matrix linear algebra) without too much
  effort.

\end{itemize}

The {\nsim} authors found that the Objective Caml system
satisfies these criteria very well. In addition, when the
implementation of the {\nsim} compiler started, there already was an
OCaml-Python interface available~\cite{Pycaml} that provided a very tight
coupling between the chosen user interface and the core implementation
language, even to the extent of making it possible to wrap up OCaml
functions to make them callable from Python and vice versa. The
implementation of {\nsim} only uses a subset of OCaml language
features which is what is known as `Core ML' plus a few extensions.
The underlying reason is that this subset of the full language on the
one hand already is sufficiently powerful to tackle practically all
problems that arouse in the implementation of {\nsim}, and on the
other hand is sufficiently simple to be mastered by students within
about three months. As an additional benefit, this voluntary
restriction greatly simplifies porting \nsim{} to Microsoft's
F\#~\cite{Fsharp}.\footnote{At the time of this writing, this is actively
  being worked on, with rather encouraging intermediate results: a
  non-parallel F\#-based port of the {\nsim} framework has been
  demonstrated in summer 2009 by M.~Franchin~\cite{nmagFsharp}.}

During the implementation of the {\nsim} prototype, the original
`Pycaml' module (which also is readily available as `pycaml' package
in the popular Debian GNU/Linux distribution) was extended
considerably by introducing new functions to simplify writing Python
extensions in OCaml. Also, a number of memory management related
problems had to be fixed (both on the Python and OCaml side). One of
the strongest benefits of this module is that it `magically' handles
all memory management related issues that have to be addressed in
order to make the OCaml garbage collector and Python reference
counting mechanism collaborate. This is a short example that
demonstrates how to extend Python with a fast (i.e. compiled to
machine code) implementation of Euler's Gamma function $\Gamma(x)$
based on extended Pycaml:

{\small
\begin{lstlisting}
=== OCaml ===

let rec euler_gamma =
  (* Accurate to eps_rel = 2e-10; See http://www.rskey.org/gamma.htm *)
  let pi = 4.0*.(atan 1.0) in
  let constants =
    [|75122.6331530; 80916.6278952; 36308.2951477;
      8687.24529705; 1168.92649479; 83.8676043424;
      2.50662827511|]
  in
  let nr_constants = Array.length constants in
    fun x ->
      if x<0.0
      then -.pi/.(x*.(euler_gamma (-.x))*.(sin (pi*.x)))
      else
	let rec compute_sum_prod sum prod pow_x n =
	  if n=nr_constants then sum/.prod
	  else 
	    compute_sum_prod
	      (sum+.constants.(n)*.pow_x)
	      (prod*.(x+.(float_of_int n)))
	      (pow_x*.x) (n+1)
	in
	let x5 = x+.5.5 in
	  exp(-.x5)*.(x5**(x+.0.5))
	  *.(compute_sum_prod 0.0 1.0 1.0 0)
in 
let _py_gamma =
  python_interfaced_function [|FloatType|]
    (fun py_args ->
       pyfloat_fromdouble (euler_gamma (pyfloat_asdouble py_args.(0))))
in
  register_for_python [|("gamma",_py_gamma)|]
;;

=== Python ===

>>> ocaml.gamma(0.5)
1.7724538509046035
>>> math.sqrt(math.pi)
1.7724538509055159
\end{lstlisting}
}

\subsection{Performance}

Comparison of the {\nsim}-based micromagnetic application {\nmag} with
another finite-element based micromagnetic package that uses very
similar technology (i.e. the same time integrator and linear algebra
libraries) allows us to get an idea of the performance of the generic
approach towards simulations relative to tailored C code. Internally,
the structure of the computations done in the main simulation loops of
{\nmag} and the finite element based micromagnetic simulator
Magpar~\cite{Magpar} are very similar, up to minor technical
details such as the question how to enforce the length constraint on
magnetization vectors and how to do residual gauge fixing. One finds
that for a given system, the time spent in the main simulation loop
for {\nmag} and Magpar usually differs by less than 20\% in either
direction. However, both setup time and memory requirements are
presently noticeably larger for {\nmag} than for Magpar. Concerning
the setup phase, this is strongly expected, as the technology
presently underlying {\nsim} resembles that of a `bytecode
interpreter' for linear operator descriptions, while Magpar uses
hand-written compiled C code to initialize its operator matrices.
Presently, the {\nsim} prototype needs about 60\% more memory for
internal bookkeeping than Magpar on simulations that are reasonably
large to be of practical relevance. There is strong evidence that this
overhead is caused by technicalities in the setup phase and can be
brought down quite considerably.

\section{Limitations}\label{sec:limitations}

The approach towards simulation physics underlying the {\nsim}
framework has limitations of different nature: fundamental limitations
that cannot be overcome, conceptual limitations where it is as yet
unclear which approach is the most promising one, as well as
technological restrictions that can (and most likely eventually will)
be overcome. As misconceptions about which problem belongs to which
category occasionally arise, it makes sense to give a short list of
the most relevant issues.

\subsection{Fundamental limitations}

Numerical simulations based on unstructured (simplicial) finite
element meshes naturally involve a considerable bookkeeping overhead
to keep track of the structure and topology of both the mesh as well
as linear operators. Hence, memory requirements are a more
constraining factor here than with finite difference based simulations
and put a stronger limit on the maximal number of degrees of freedom
that can be included in a simulation. While much of the underlying
structure of {\nsim} could be extended rather easily to support finite
differences and structured meshes in addition to finite elements, the
fundamental limitation of increased memory requirements for
unstructured finite element based approaches cannot be overcome.

\subsection{Conceptual limitations}

While most of the pieces for a simple unified description language for
physical systems have been implemented, this has not yet been
integrated in such a way that a parallel simulation can be produced
from a single textual description file. Rather, the specification has
to be done piece-wise for individual parts such as differential
operators, local field operations, etc. Technically speaking, this can
be easily overcome by combining these pieces into a unified LALR(1)
grammar. However, some unresolved conceptual issues include finding a
suitable (i.e. convenient) notation for computations which presently
have to be specified as sequences of linear algebra operations (such
as for the computation of the demagnetizing field in micromagnetism).

By implementing both support for key techniques in conjunction with
convenient notational conventions, a simulation framework can gain
considerable expressive power and broad applicability. From the
perspective of renormalization theory, we would expect situations to
be rare where a field theoretic model has to employ concepts beyond
these:

\begin{itemize}

\item Differential operators of order 1 and 2.

\item Local (i.e. nondifferential) non-linear operators.

\item Fields mediated by very light or massless particles which give rise to 
`Coulomb's Law' type interactions.
\end{itemize}

However, this perspective leaves open many important questions on what
approach to discretisation will behave how well or how badly.

\subsection{Technical restrictions of the \nsim{} prototype}

For the purpose of providing an overview, we summarize the most
relevant open issues that at the time of this writing have not been 
implemented in the public release of the {\nsim} prototype. 
Most of these have been described in more detail in the preceding text.

\begin{itemize}

\item The {\tt fastfields} dynamic linking mechanism at present 
easily gets confused if it is ever fed broken C code. Ideally, calling
a dynamically compiled function for which a broken definition was given
should throw an exception which contains both the compiler error
message and some context from the code provided by the user. This is
achievable, but has not been implemented yet.

\item Error reporting can (and eventually must) be improved considerably.
 At present, there are still a few situations where a minor mistake in
 a simulation specification can produce a highly cryptic error message
 (such as ``index out of bounds'') that is hard to track down.

\item Support for dynamic rebuilding of sparse and nonsparse 
operator matrices as well as the inclusion of `middle fields' still is
fragmentary in the public release.

\item At present, field dependencies must be specified by the user. 
While it would be rather straightforward to auto-derive the field
dependencies from the specifications of the linear algebra execution
sequences as well as the user-specified local equations, this has not
been implemented yet.

\item Startup argument handling of {\nsim} still is rather ad-hoc. 
Basically, there are six different kinds of program arguments that have
to be dealt with:
\begin{itemize}
\item MPI startup arguments
\item PETSc related arguments
\item Arguments controlling the Python interpreter
\item {\nsim} arguments (controlling e.g. the nsim loglevel)
\item {\nsim} library arguments (e.g. when using the {\nmag} library)
\item Arguments to the user's application/script.
\end{itemize}
The {\nsim} team has a strategy for eventually handling this
complexity in a convenient way, but this has not been implemented yet.

\item An important limitation of the extended Pycaml Python-OCaml 
interface module (to be released separately as a spin-off of the
{\nsim} project) is that it is not thread-safe with respect to Python
threads. While this has no relevance to the {\nsim} project at
present, the wider community may nevertheless benefit from making this
module thread-safe.

\item A major re-structuring of the code that auto-generates the
  Jacobian from the equations of motion has not been fully completed
  yet.

\item {\nsim} does not support complex-valued fields (and operators), as 
the PETSc library cannot handle both complex and real vectors at the
same time. In order to resolve this, one would most likely have to
wait for a major change to the PETSc library or find a way to compile
it in such a way that two variants (which use different symbol
prefixes) can be linked at the same time, one for real valued
computations, the other one for complex computations (which then would
most likely use different MPI communicators).

\item {\nsim} presently neither supports dynamic mesh refinement nor
  non-simplicial mesh geometries. (Considering refinement, this is
  considered to be difficult to get right in micromagnetism, avoiding
  numerical artefacts.)

\end{itemize}

\section{Conclusion and outlook}\label{sec:conclusion-outlook}

The viability of automated compilation of symbolically specified
physical simulations to fast parallel code in a way that preserves
user-level sequential computing semantics has been demonstrated
through {\nsim}. Furthermore, the practicability of such an approach
as a basis for simulations of scientific relevance also has been
demonstrated through the micromagnetism library {\nmag} implemented on
top of {\nsim}. As practicality and usability are the main driving
forces underlying the design of the {\nsim} framework, many strategic
decisions were relentlessly revamped (sometimes multiple times)
whenever the authors discovered that an early design decision had lead
to more inconvenience in usage than was originally expected. As such,
much of the design of {\nsim} must still be considered a `moving
target' that has not stabilized in its final form yet. But for
precisely this reason, malleability of the higher levels close to the
end user, it is considered important to inform the wider academic
community of the status of this project at the present state and
utilize the academic process to guide further design decisions -- now
that the structure of the system's simulation core has stabilized --
in those directions most useful for applications.

\bigbreak
{\bf Acknowledgments\hfill}
\smallskip
\noindent It is a pleasure to thank numerous people who have influenced the
development of the {\nsim} framework in a variety of ways. 

In particular, we want to thank Werner Scholz for providing us with
both strategic advice on issues related to time integration as well as
for writing Magpar, which was very instructive for the design of
{\nmag}. We furthermore thank Matteo Franchin for implementing a
number of key components of the software framework (including the
platform-independent installation system), Giuliano Bordignon for his
work on meshing related code and algorithms, Jacek Generowicz for
strategic help with Python and with the documentation, Michael Walter
for finishing both the implementation of Langevin dynamics as well as
support for local polynomial equations, Dmitry Grebeniuk for allowing
us to include his OCaml memory footprint code in {\nsim}. Finally,
T.F.  wants to thank his wife Helena Fischbacher-Weitz for moral
support during this voluminous project. This work was supported by
EPSRC grants GR/T09156/01 and EP/E040063/1. The research leading to
these results has received funding from the European Community's
Seventh Framework Programme (FP7/2007-2013) under Grant Agreement
233552.

\appendix

\section{A complete {\nsim} example}\label{sec:appendix}

An example for level~3 Python user code on top of the {\nmag}
micromagnetic library has been given in
figure~\ref{fig:micromag-example-code}. This appendix provides a
complementary level~2 Python example that shows how to employ the
concepts described in the main body of this work to set up and
simulate a complete physical model that involves PDEs. Also, it shows
how to combine a {\nsim} simulation with other software libraries for
setup, data analysis, and plotting. While the simulation script is
essentially a sequential program, it can be started in parallel on a
cluster and will then utilize MPI-based linear algebra to distribute
the simulation-related computational effort across nodes. The data
analysis part always gets executed as sequential code on the master
node.

\subsection{The physical system}

The system to be studied is a reaction-diffusion system that was
originally proposed by Alan Turing as a model for morphogenesis in
biological systems~\cite{Turing1952}. While the framework provides all
the bookkeeping infrastructure (as explained in the main part of this
work), the level~2 interface of {\nsim} is still somewhat rudimentary
at the time of this writing, and could be simplified considerably by
adding more inference logic and introducing a dedicated grammar that
can be used to express a complete task specification: with this, it
should be possible to bring down the programming effort required to
specify the physics of a field theoretic system like the one presented
here from presently about 75~lines to less than~20. One reason for the
present redundancy is that, during {\nsim}'s evolution, major design
changes have occurred, such as the replacement of the original
parallel timestepper functionality with new code that is more tightly
integrated with the main part of the simulation. Such low-level
structural changes favor some independence in higher-level
specifications, which here shows up as redundancy.

The system under study can arise as a linearization around a
stationary point of a binary chemical system whose position-dependent
concentrations~$u(x,y),\,v(x,y)$ change according to:
\begin{equation}
\begin{array}{lcl}
\frac{d}{dt} u&=& f_u(u,v) + D_u\Delta u\\
\frac{d}{dt} v&=& f_v(u,v) + D_v\Delta v.
\end{array}
\end{equation}

Let us assume that the truncated system that is obtained by setting
the diffusion constants~$D_u=D_v=0$ has a stationary point at
$(u_0,v_0)$. Then, the linearized system that is obtained by taking
the $\epsilon\to0$ limit of $u=u_0+\epsilon\tilde u$,
$v=v_0+\epsilon\tilde v$ is governed by the equation
\begin{equation}
\frac{d}{dt}\left(\begin{array}{c}\tilde u\\\tilde v\end{array}\right)=
\left(\begin{array}{cc}
\frac{\partial f_u}{\partial u}&\frac{\partial f_u}{\partial v}\\
\frac{\partial f_v}{\partial u}&\frac{\partial f_v}{\partial v}
\end{array}\right)_{|(u_0,v_0)}\cdot
\left(\begin{array}{c}\tilde u\\\tilde v\end{array}\right)
+\left(\begin{array}{c}D_u\Delta\tilde u\\D_v\Delta\tilde v\end{array}\right).
\end{equation}

\begin{figure}
  \centering
  \begin{tabular}{cc}
  \includegraphics[width=6cm]{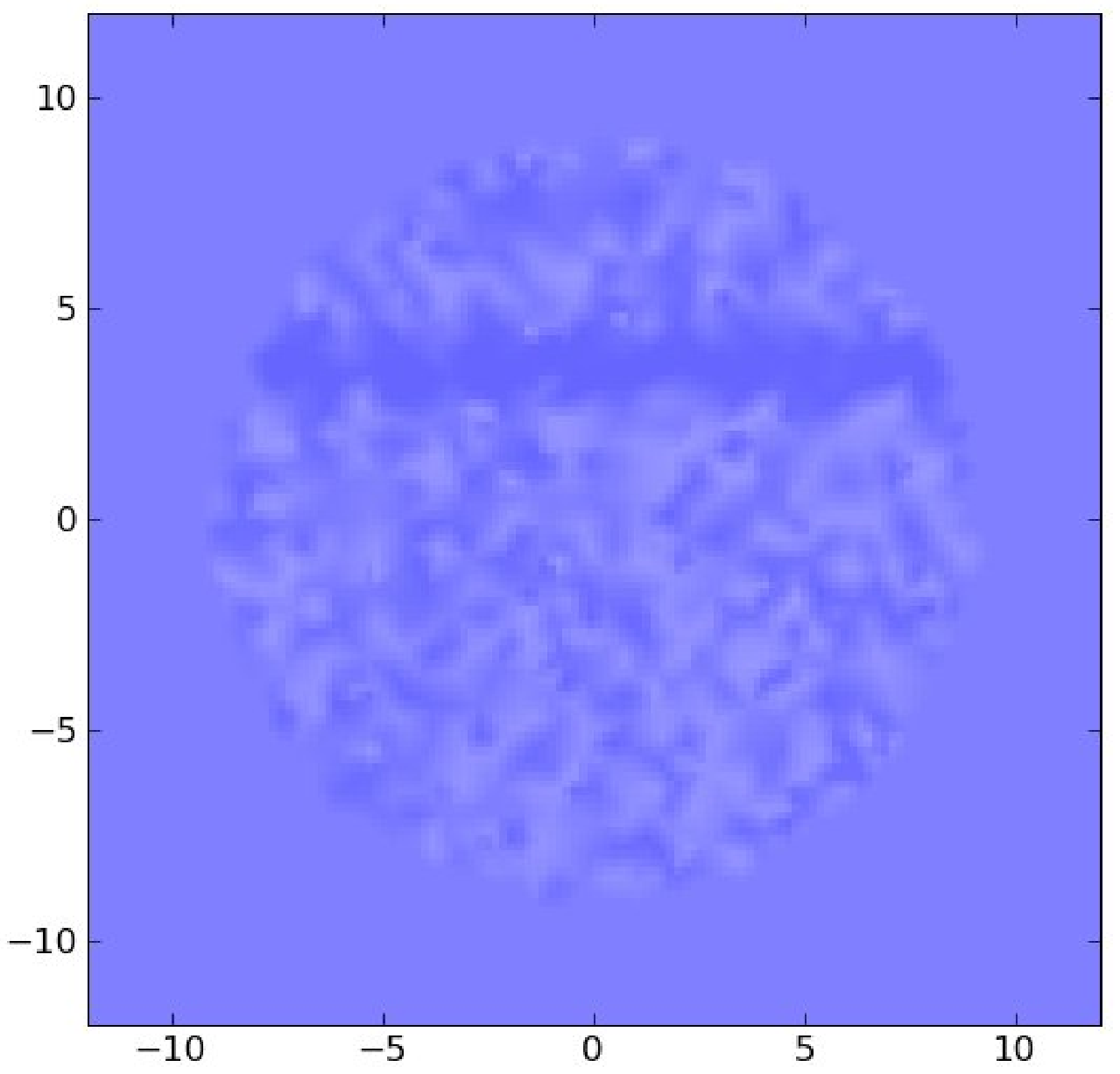}&
  \includegraphics[width=6cm]{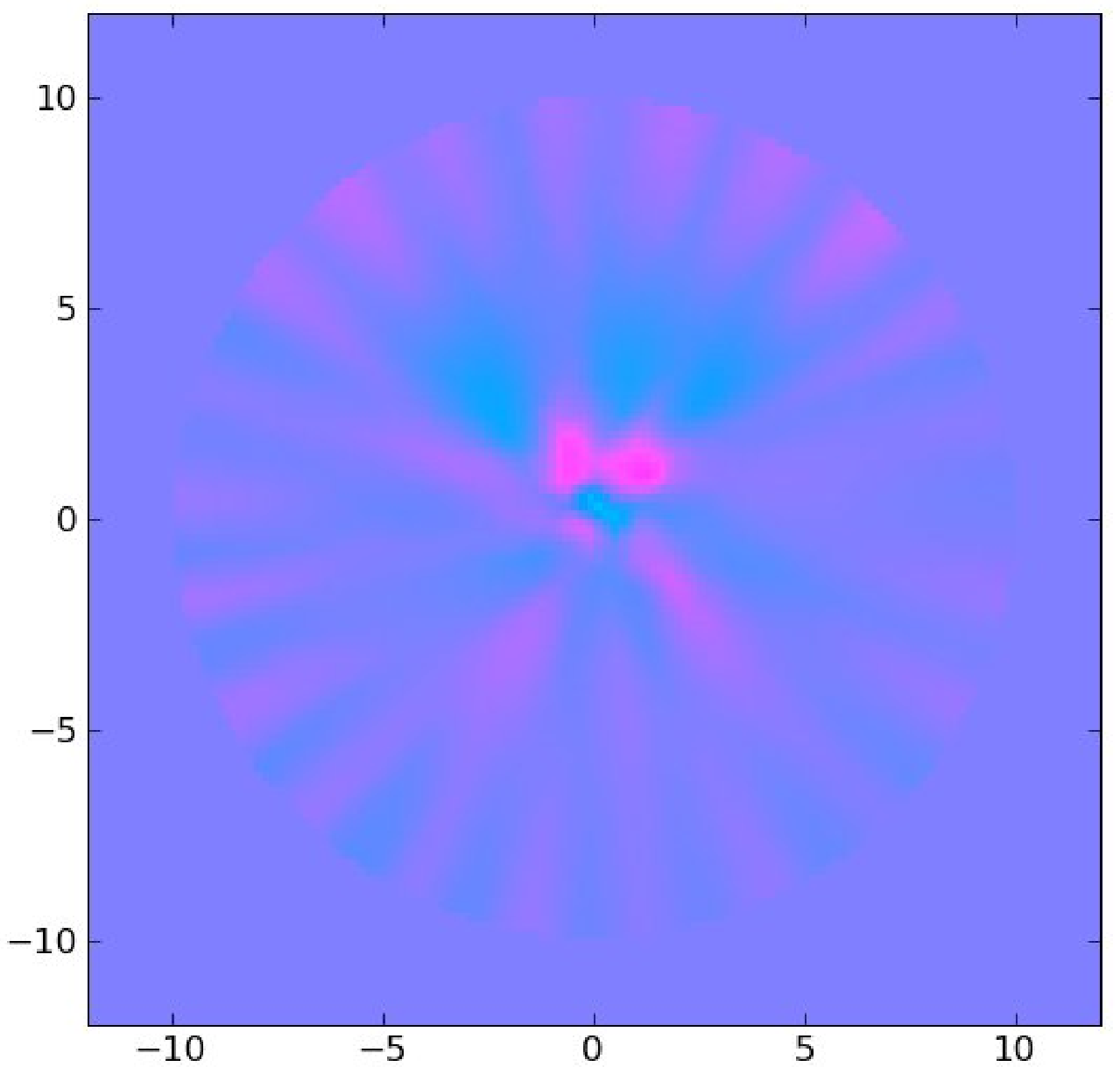}\\
  \includegraphics[width=6cm]{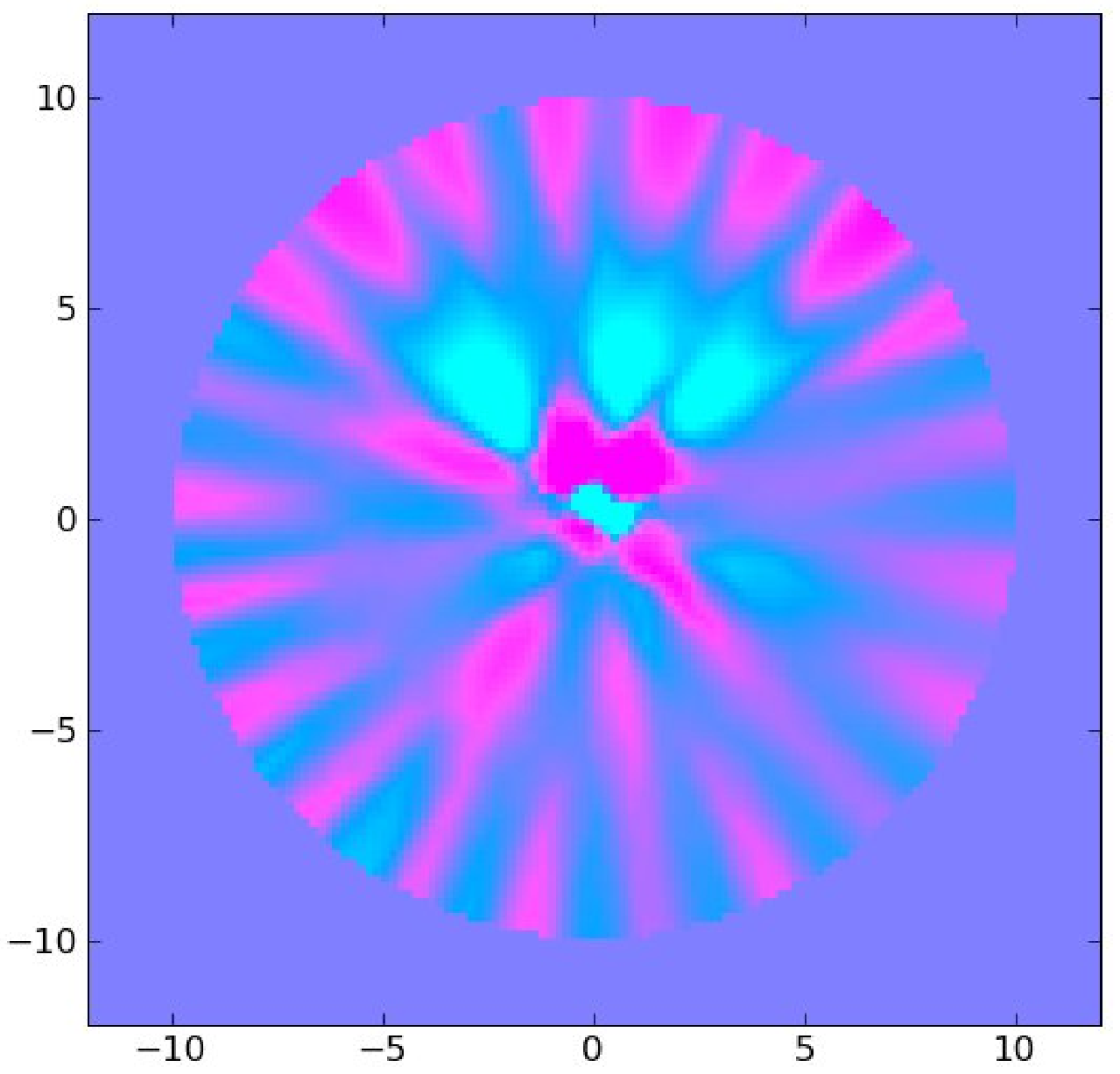}&
  \includegraphics[width=6cm]{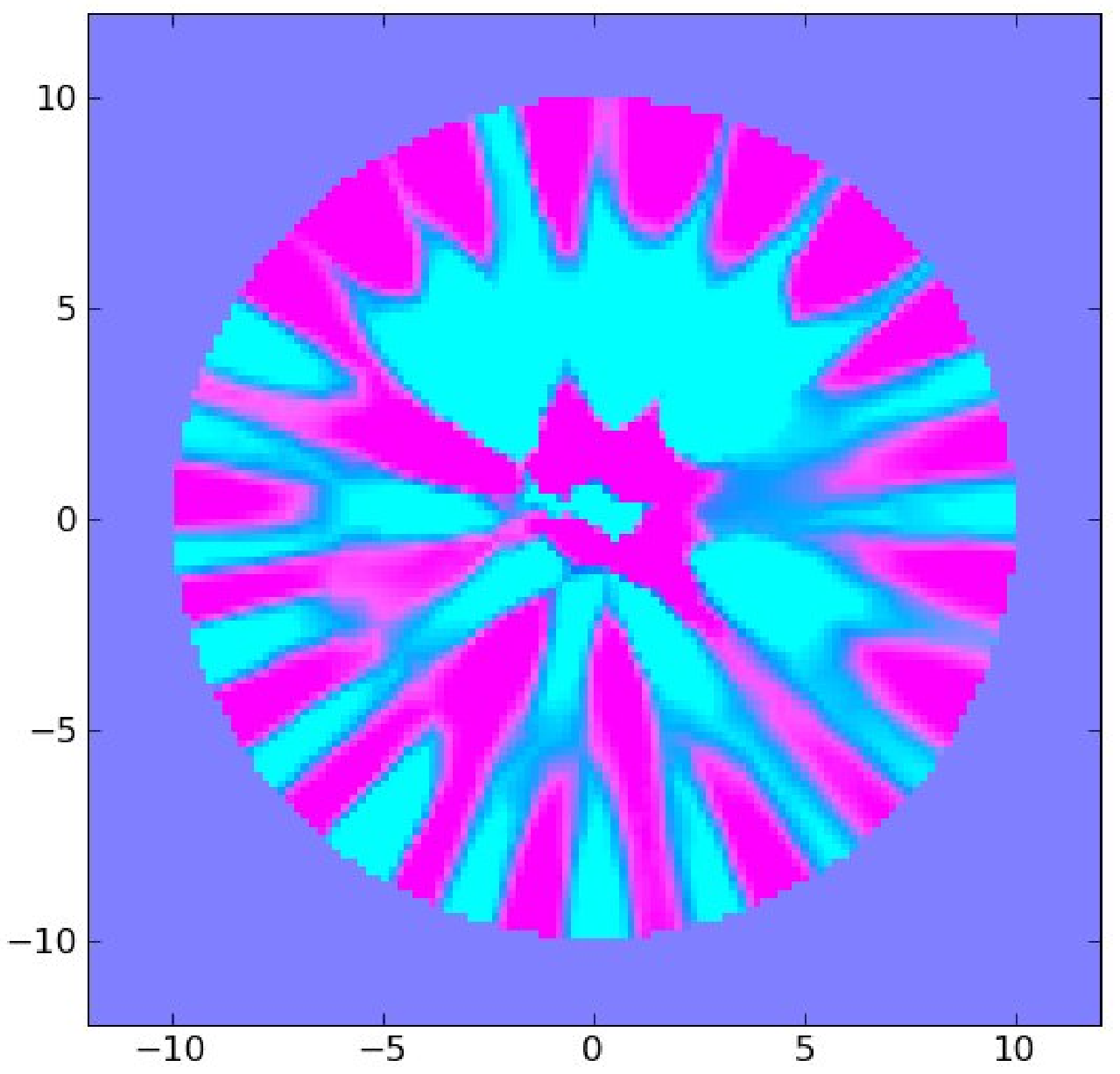}
  \end{tabular}
  \caption{Pattern formation from random initial noise by
    the Turing instability in a reaction-diffusion system.
    \label{fig:turingpatterns}}
\end{figure}

If a number of conditions on the diffusion constants and the matrix of
partial derivatives are satisfied, then the situation can arise that
the system's stability with respect to spatially oscillating
fluctuations of the form
\begin{equation}
(\tilde u(\vec x),\tilde v(\vec x)) = (\tilde u_*,\tilde v_*)\cdot\cos(\vec k\cdot\vec x)
\end{equation}
is violated for some range of wave-vectors~$\vec k$ 
with~$k_{\rm min}<|\vec k|<k_{\rm max}$. Over this range of wave-vectors,
initial random fluctuations get amplified, which leads to the formation
of patterns, as shown in figure~\ref{fig:turingpatterns}.

The parameters chosen for this example are:

\begin{equation}
\left(\begin{array}{cc}
\frac{\partial f_u}{\partial u}&\frac{\partial f_u}{\partial v}\\
\frac{\partial f_v}{\partial u}&\frac{\partial f_v}{\partial v}
\end{array}\right)=
\left(\begin{array}{cc}
+1&+2\\
-2&-3
\end{array}\right),\qquad D_u=-0.025,\quad D_v=-0.25
\end{equation}
on a disk with radius $R=10$. Note that the trace of the linearization
matrix is negative, while the determinant is positive, so the
realparts of the eigenvalues are negative, and the system is stable in
the long wavelength ($k\rightarrow0$) limit. For intermediate values 
of $|\vec k|$, the negative contribution to the determinant from the 
diagonal entries over-compensates the positive off-diagonal contribution 
to the determinant, so the determinant becomes positive while the trace
remains negative, i.e. one of the two eigenvalues will obtain a positive 
real part, which causes the instability. For large~$|k|$, the diagonal 
again gives a large positive contribution, and the realparts of the 
eigenvalues are negative again.

\subsection{The Python code}

The complete Python program that produced this output is given in the
following. The central part that sets up the physics simulation engine
is the call to \texttt{make\_linalg\_machine()}. This translates a
full specification of field layouts, linear operators, local reaction
equations, short sequences of elementary parallelizable operations,
and numerical tweaking parameters to an opaque object that can be used
to perform time integration in such a system. The part of the code
starting with the \texttt{sample\_field()} function deals with data
extraction and visualization only.

{\small
\begin{lstlisting}
import math, numpy
from nsim.linalg_machine import *

# Work around a memory management problem of pylab:
import matplotlib
matplotlib.use('Agg')
import pylab

(R_disk,s_disk) = (10.0, 0.5)
n_boundary=int(1+2*math.pi*R_disk/s_disk)
n_interior=int(1+math.pi*R_disk**2/(s_disk**2))

disk_boundary=[[R_disk*math.cos(2*math.pi*n/n_boundary),
                R_disk*math.sin(2*math.pi*n/n_boundary)]
               for n in range(n_boundary)]

# Generate a mesh - this will just have one material region,
# whose number is 1. (This mesh-generating function gives fairly
# poor results and is supposed to be used for testing purposes only!)
mesh_disk=ocaml.simple_convex_mesh(disk_boundary,0,n_interior,10)

# Generate a postscript file that shows the 2d mesh:
ocaml.mesh2d_ps(mesh_disk,"/tmp/disk%d.ps"%n,[10.0,10.0,200.0,200.0])

(elem_dim,elem_order)=(2,1)
elem_uv = ocaml.fuse_elements(\
    ocaml.make_element("u", [], elem_dim, elem_order),
    ocaml.make_element("v", [], elem_dim, elem_order))

mwe_uv = ocaml.make_mwe("uv",mesh_disk,
                        [(0,ocaml.empty_element), # "vacuum"
                         (1,elem_uv)],[],[])

mwe_uv_dot = ocaml.mwe_sibling(mwe_uv,"uv_dot",
                               "d_dt_", # element renaming prefix
                               [("u","u_dot"), ("v","v_dot")])

mwe_uv_diff = ocaml.mwe_sibling(mwe_uv,"uv_diff",
                                "Diffusion_",
                                [("u","u_diff"), ("v","v_diff")])

# Equations of motion:
eq_ydot = """
u_dot <- u_diff + (+1.0)*u + (+2.0)*v;
v_dot <- v_diff + (-2.0)*u + (-3.0)*v;"""

the_lam=make_linalg_machine("Turing",
    mwes=[mwe_uv,mwe_uv_dot,mwe_uv_diff],
    vectors=[lam_vector(name="v_uv",mwe_name="uv"),
             lam_vector(name="v_uv_dot",mwe_name="uv_dot"),
             lam_vector(name="v_uv_diff",mwe_name="uv_diff")],
    operators=[\
      lam_operator("op_diff","uv_diff","uv",\
        "-0.025*<d/dxj u_diff||d/dxj u>\
         +(-0.25)*<d/dxj v_diff||d/dxj v>, j:3")],
    local_operations=[\
        lam_local("local_uv_dot",
                  field_mwes=["uv_dot","uv_diff","uv"],
                  equation=eq_ydot)],
    programs=[\
        lam_program("set_uv_diff",
            commands=[["SM*V","op_diff","v_uv","v_uv_diff"],
                      ["CFBOX","uv_diff","v_uv_diff"]]),
        lam_program("update_uv_dot",
            commands=[["GOSUB","set_uv_diff"],
                      ["SITE-WISE-IPARAMS","local_uv_dot",
                       ["v_uv_dot","v_uv_diff","v_uv"],[]]]),
            ],
    timesteppers=[
        lam_timestepper("TS",\
            ["uv","uv_dot","uv_diff"],
            ["v_uv","v_uv_dot","v_uv_diff"],
            "update_uv_dot",
            nr_primary_fields=1,
            pc_rtol=1e-8, pc_atol=1e-8,
            max_order=2, krylov_max=5,
            jacobi_eom=eq_ydot,
            phys_field_derivs=[("PRIMARY",""),
                               ("PRIMARY",""),
                               ("OPERATOR","op_diff")],
            jacobi_prealloc_diagonal=50,
            jacobi_prealloc_off_diagonal=50)])

def advance_time(field,times,do=None):
    t_initial=times[0]

    ocaml.linalg_machine_set_field(the_lam,field, "v_uv")
    ocaml.linalg_machine_initialize_timestepper(\
        the_lam,"TS", t_initial,
        1e-7,1e-7) # rel and abs tolerances
    n=0
    if do: do(field_uv,t_initial,n)
    for time in times[1:]:
        ocaml.linalg_machine_advance_timestepper(\
            the_lam, "TS", time,
            1000000) # max iterations
        n=n+1
        ocaml.linalg_machine_get_field(the_lam,field,"v_uv")
        if do: do(field_uv,time,n)
    return field

def sample_field(field,radius,npoints=128):
    v=numpy.zeros([npoints,npoints,2])
    step=2*radius/(npoints-1.0)
    min_coord=-step*(npoints-1.0)/2.0
    coord_range=[min_coord+n*step for n in range(npoints)]
    for yc in range(npoints):
        for xc in range(npoints):
            pos=[coord_range[xc],coord_range[yc]]
            z=ocaml.probe_field(field,"u",pos)
            # z is e.g. [('u', 37.511166751728595)], or []
            if len(z)>0:
                v[yc,xc,0]=z[0][1]
                z=ocaml.probe_field(field,"v",pos)
                v[yc,xc,1]=z[0][1]
    return (v,-min_coord)

def plot_sampled(sampled,color_range=1.0):
    pylab.hold(False)
    for nr_plot in range(len(sampled)):
        print "Doing plot:",nr_plot
        (data,xmax)=sampled[nr_plot]
        npoints=data.shape[0]
        cdata=numpy.zeros([npoints,npoints,4],dtype=numpy.uint8)
        for yc in range(npoints):
            for xc in range(npoints):
                u=data[yc,xc,0]
                v=data[yc,xc,1]
                red=int(255*(0.5+u/color_range))
                green=int(255*(0.5+v/color_range))
                red=max(0,min(red,255))
                green=max(0,min(green,255))
                cdata[yc,xc]=(red,green,255,255)
                pylab.imshow(cdata,
                             interpolation="nearest",
                             extent=(-xmax,xmax,-xmax,xmax))
        pylab.savefig("/tmp/turing%02d.png"%nr_plot)

def fun_uv(dof,pos):
    """This function produces some initial all-wavelength 'noise'"""
    (dof_stem,dof_indices)=dof
    if pos[0]*pos[0]+pos[1]*pos[1]>80.0:
        return 0.0 # No fluctuations close to the boundary
    # Pseudo-Random fluctuations otherwise:
    return (((((4*math.pi+pos[0])**8)%1.0+\
              math.pi+pos[1])**8)%1.0-0.5)*1e-3

field_uv=ocaml.raw_make_field(\
    mwe_uv, [fun_uv], # opt. sampling fun
    "","")

plots=[]

def do_add_plot(field,time,n):
    print "\n*** DOING T=%.3f ***\n"%time
    plots.append(sample_field(field,1.2*R_disk))

advance_time(field_uv,[x*50.0 for x in range(4)],
             do=do_add_plot)

plot_sampled(plots,color_range=5e-3)
\end{lstlisting}
}

\end{document}